\documentclass[fleqn,usenatbib]{mnras}

\usepackage{newtxtext,newtxmath}

\usepackage[T1]{fontenc}
\DeclareRobustCommand{\VAN}[3]{#2}
\let\VANthebibliography\thebibliography
\def\thebibliography{\DeclareRobustCommand{\VAN}[3]{##3}\VANthebibliography}

\usepackage{graphicx}	
\usepackage{amsmath}	
\usepackage[dvipsnames]{xcolor} 
\usepackage[flushleft]{threeparttable} 
\usepackage{float}
\usepackage{booktabs} 
\usepackage{physics} 
\usepackage{subcaption}
\usepackage[normalem]{ulem}
\usepackage{caption}

\title[]{Action-based dynamical models of M31-like galaxies}

\author[P. Gherghinescu et al.]{
Paula Gherghinescu$^{1}$,\thanks{E-mail: p.gherghinescu@surrey.ac.uk}
Payel Das$^{1}$, Robert J. J. Grand$^{2}$, Matthew D. A. Orkney$^{3,4}$ \\
$^{1}$Physics Department, University of Surrey, Guildford GU2 7XH, United Kingdom \\
$^{2}$Astrophsics Research Institute, Liverpool John Moores University, 146 Brownlow Hill, Liverpool L3 5RF \\
$^{3}$Institut de Ci\`{e}ncies del Cosmos (ICCUB), Universitat de Barcelona, Mart\'{i} i Franqu\`{e}s 1, E-08028 Barcelona, Spain\\
$^{4}$Institut d'Estudis Espacials de Catalunya (IEEC), E-08034 Barcelona, Spain
}

\date{Accepted XXX. Received YYY; in original form ZZZ}

\pubyear{2022}

\begin{document}
\label{firstpage}
\pagerange{\pageref{firstpage}--\pageref{lastpage}}
\maketitle

\begin{abstract}
In this work, we present an action-based dynamical equilibrium model to constrain the phase-space distribution of stars in the stellar halo, present-day dark matter distribution, and the total mass distribution in M31-like galaxies. The model comprises a three-component gravitational potential (stellar bulge, stellar disk, and a dark matter halo), and a double-power law distribution function (DF), $f(\mathbf{J})$, which is a function of actions. A Bayesian model-fitting algorithm was implemented that enabled both parameters of the potential and DF to be explored.

After testing the model-fitting algorithm on mock data drawn from the model itself, it was applied to a set of three M31-like haloes from the Auriga simulations (Auriga 21, Auriga 23, Auriga 24). Furthermore, we tested the equilibrium assumption and the ability of a double-power law distribution function to represent the stellar halo stars. The model incurs an error in the total enclosed mass of around 10 percent out to 100 kpc, thus justifying the equilibrium assumption. Furthermore, the double-power law DF used proves to be an appropriate description of the investigated M31-like halos. The anisotropy profiles of the halos were also investigated and discussed from a merger history point of view. 
\end{abstract}





\section{Introduction}

Stellar haloes play an important role in understanding the accretion histories of galaxies. \citet{Bell+2008} found that the majority of the stellar halo in the Milky Way (MW) is primarily composed of substructure, originating from external, accreted galaxies through stripping, \citep[e.g.,][]{Johnston+1996, Johnston1998, Read+2006, Naidu+2020}, but also stars from the inner parts of the galaxies, that have been heated into wider orbits \citep[e.g][]{Zolotov+2009,Cooper+2015}. Due to the long timescales of energy and momenta exchange compared to the ages of the host galaxies, we can expect to find preserved relics of past accretion events in phase space. Even when these substructures are phase mixed, there is a memory of the original accreted satellites in the chemical composition and dynamical properties of the stars. Moreover, the galactic halo is dark matter dominated with stellar content being low. This makes stellar haloes invaluable in investigating the dark matter content of galaxies.
 
\par Thanks to state-of-the-art instruments, such as those aboard Gaia \citep{GaiaPaper2016}, we now have the biggest, most accurate 6D phase-space map of our own Galaxy. However, our position inside the MW can hinder the study of the global halo. Furthermore, our Galaxy is just one of the many disk spiral galaxies in the Universe. If we wish to get a complete picture of the pathways leading to the formation of disk galaxies, we need to expand our sample beyond the MW. In this regard, studying our neighbor, the Andromeda Galaxy (M31) is complementary. Its proximity ( $\sim 780$ kpc) and edge-on orientation ($i\sim 77^\circ$) offer a panoramic view and make it an ideal candidate for studying stellar haloes of external disk galaxies.
 
\par The stellar halo of M31 shows a wealth of substructures, as well as numerous surviving satellites \citep[e.g][]{ Martin+2009, Richardson+2011, McConnachie+2018_PandasLSSII}, suggesting a busy recent accretion history. There is evidence of a major recent accretion event that happened $<4\,\text{Gyr}$ ago with an external galaxy of mass $\sim 10^{10}M_{\odot}$ \citep[e.g.,][]{ Hammer+2018, Bhattacharya+2019_PNeII}. In contrast, MW's evolution appears to have been more secular in the last $\approx 8-9$ Gyr. There is growing evidence that the MW experienced a significant minor merger $\approx 8$ Gyr ago but nothing since \citep[e.g.,][]{Helmi+2018, Belokurov+2018}.

\par The density profile of the stellar halo of the MW exhibits a break around $15$ to $25\,\text{kpc}$ \citep[e.g.][]{Sesar+2011, Deason+2011, Kafle+2014}. In contrast, the density profile of the stellar halo of M31 \citep[e.g.][]{Gilbert+2012} shows a smooth profile, with no break, out to 100 kpc. \cite{Deason+2013} used the simulations of \cite{Bullock+Johnston2005} to associate the break in the MW density profile to a relatively early and massive accretion event (likely Gaia-Enceladus Sausage). In contrast, the lack of a break in the M31 stellar halo profile suggests that its accreted satellites have a wide range of apocenters, over a more prolonged accretion history.

\par Through surveys such as SPLASH (Spectroscopic and Photometric Landscape of Andromeda’s Stellar Halo, using the DEIMOS instrument on the Keck II telescope, \citealp{SPLASH2012}) and PAndAS (The Pan-Andromeda Archaeological Survey, collected with the MegaCam wide-field camera the Canada France Hawaii Telescope,  \citealp{McConnachie+2009_PandasSurvey}), photometry and spectroscopy of individually resolved stars in the disk and halo of M31 have been obtained. Despite the low surface brightness of M31 beyond the disk, its stellar halo can be investigated using discrete tracers. Globular clusters have been successful in tracing the outer halo ($>50$ kpc) substructures \citep[e.g.,][]{Mackey+2010_GCM31, Velajanoski+2014_GCM31} while Planetary Nebulae have been used in investigating the inner parts of the halo \citep[e.g.,][]{Bhattacharya+2019_PNeM31}.

One path to understanding the history of a galaxy is by looking at its present-day dark matter distribution and characterising the phase-space distribution of stars in its stellar halo. Under the assumption of dynamical equilibrium, the two are directly connected. In reality, the equilibrium assumption is likely only approximate (e.g., due to recent interactions with other galaxies), but nonetheless very useful. Plenty of studies have made use of this assumption to extract mass profiles of galaxies \citep[e.g.,][]{Thomas+07, Das+2011, Piffl+14, Portail+17, Vasiliev2019, zhu+23}. 

\par One approach to specifying dynamical equilibrium models is through distribution functions (DFs). The DF can be interpreted as the probability of finding a star in an infinitesimally small phase-space volume. The DF stores, in a single functional form, information about a wide range of properties, such as radial profiles of density, flattening, rotation, and velocity anisotropy of a stellar system. We can most simply construct DFs through functions of constants of motion \citep{Jeans1915}. There are clear advantages in taking these constants of motion to be the actions.

\par The angle-action variables ($\vb*{\theta,J}$), are obtained through a canonical transformation of the (\textit{\textbf{x, v}}) coordinates. Working with these variables has many advantages: actions are smooth functions which makes them more straightforward to work with in mathematical models. They are adiabatically invariant, making them natural variables to use for perturbation theory. Furthermore, the actions, $\vb*{J}$, have a straightforward interpretation: the radial action $J_{r}$ measures the eccentricity of an orbit, the vertical action $J_{z}$ describes the wandering beyond the galactic plane, while the azimuthal action $J_{\phi}$ quantifies the degree of prograde or retrograde motion \citep{Binney&Tremaine1987_GalDynamics}.  

\par Double-power law action-based DFs are flexible and versatile in modeling stellar haloes of galaxies, allowing the model to capture varying behviour of the stellar halo at small and large radii. These DFs have already been proven to fit a wide range of observed properties in the MW \citep[e.g][]{Piffl+14, Das+Binney2016_EDF1, Das+2016_EDF2, Hattori+2021}, but in this work, we investigate their suitability for M31-like haloes. In this paper, we fit the double-power law, action-based DFs to M31-like stellar haloes from the Auriga simulations \citep{Grand+2017_AurigaSims}. This will both allow us to test the appropriateness of double-power law DFs as a description for M31-like haloes, and the ability of the model, in the context of the dynamical equilibrium assumption, to recover the total mass profile and dark matter content. We will also use the simulations to investigate the connection between the accretion history and the velocity anisotropy profiles.

\par This paper is organised as follows: in Section \ref{sec:method}, we present the equilibrium dynamical model and the Bayesian-fitting pipeline. In Section \ref{sec:mock_data}, we test the pipeline on a mock galaxy. In Section \ref{sec:auriga} we apply the pipeline to three M31-like Auriga haloes. We interpret and discuss our findings in Section \ref{sec:discussion}, and conclude in Section \ref{sec:conclusion}.

\section{A dynamical equilibrium model for the stellar halo} \label{sec:method}

In this Section, we present the gravitational potential and double power-law action-based DF that will be used to model the stellar halo of M31, and the procedure used to fit for their parameters. The model is implemented using the \texttt{AGAMA} package \citep{Vasiliev2019_Agama}. \texttt{AGAMA} (action-based galaxy modelling architecture) is a software library that provides tools for galaxy modelling. The package contains mathematical routines and frameworks for constructing models through: gravitational potential objects, transformation between position/velocity to angle/action space, DF objects expressed in terms of actions, etc.

\subsection{The gravitational potential} \label{Method:potential}
The gravitational potential is assumed to be an oblate axisymmetric composite potential comprising a bulge, disk, and dark matter halo. Density profiles are defined for each of these components from which the potential is computed through the Poisson equation. The density of the disk is given by:
\begin{equation}
    \rho_{\mathrm{d}}(R,z) = \Sigma_{0,\text{d}} \exp[-\biggl( \frac{R}{R_{\text{d}}}\biggr)^{\frac{1}{n_{\text{d}}}}]\times \frac{1}{2h} \exp(-\abs{\frac{z}{h}}),
\end{equation}
where $R = \sqrt{x^2 + y^2}$ with $(x,y)$ the Cartesian coordinates of the star in the plane of the disk, $\Sigma_{0,\text{d}}$ is the surface density (but can be specified through the total mass of the component), $R_{\text{d}}$ is the scale radius, $h$ is the scale height of the disk (i.e., a measure of the vertical extension), and $n_{\text{d}}$ is the Sérsic index.

The density for the bulge is obtained through the (deprojected) Sérsic surface density distribution:
\begin{equation}
\Sigma_{\text{b}}(R) = \Sigma_{0,\text{b}}\exp[-b_{n}\biggl(\frac{R}{R_{\text{b}}} \biggr)^\frac{1}{n_{\text{b}}}],
\end{equation}
with $R_{\text{b}}$ being the bulge scale radius and $b_{n}\approx 2n_{\text{b}}-1/3$, where $n_{\text{b}}$ is the Sérsic index. $R$ is as above, $\Sigma_{0,\text{b}}$ is the surface density, and $R_{\text{b}}$ is the scale radius.

\par Finally, we have chosen an NFW profile \citep{Navarro+1997_NFW} for the dark matter density profile:
\begin{equation}
    \rho_{\text{dm}}(R,z) = \frac{\rho_{0,\text{dm}}}{\frac{r}{R_{s}}\left(1+\frac{r}{R_{s}}\right)^{2}},
\end{equation}
with $\rho_{0,\text{dm}}$ the normalisation density and $R_{s}$ the scale radius. The radius $r = \sqrt{x^2+y^2+(z/q)^2} = \sqrt{R^2+(z/q)^2}$ takes into account the flattening of the halo through the $q$ parameter, which measures the flattening of the halo along the $z$-direction. Therefore, the assumed DM halo is oblate axisymmetric. As potentials are additive, the total potential of the galaxy is the sum of the potentials generated by the individual components,
\begin{equation}
    \Phi(R,z) = \Phi_{\mathrm{d}}(R,z) + \Phi_{\mathrm{b}}(R,z) + \Phi_{\mathrm{DM}}(R,z),
\end{equation}
where $\Phi_{\mathrm{d}}(R,z), \Phi_{\mathrm{b}}(R,z),$ and $\Phi_{\mathrm{DM}}(R,z)$ are the potentials of the disk, bulge, and dark matter halo, respectively.

\subsection{The stellar halo DF} \label{Method:dfs}
To model the action-space distribution of stars in the stellar halo, we use a generalisation of the DF proposed by \citet{Posti+2015_sphDF} for spheroidal galactic components. This generalised DF has been implemented in \texttt{AGAMA} and has the following functional form:

\begin{align} 
\begin{split}
   & f(\vb*{J})  = \frac{M_{0}}{(2\pi J_{0})^{3}}\biggl[ 1+\frac{J_{0}}{h(\vb*{J})}\biggr]^{\alpha} \biggl[ 1+\frac{g(\vb*{J})}{J_{0}}\biggr]^{(\alpha-\beta)} \biggl(1+\chi \tanh\frac{J_{\phi}}{J_{\phi,0}}\biggr), \\ 
& \text{where}\,\,g(\vb*{J}) = g_{r}J_{r} + g_{z}J_{z} + (3- g_{r}-g_{z})\left | J_{\phi} \right |, \\
& \text{and}\,\,h(\vb*{J}) = h_{r}J_{r} + h_{z}J_{z} + (3- h_{r}-h_{z})\left | J_{\phi} \right |.
\end{split}
\end{align}

$M_{0}$ is a normalization parameter with units of mass (which differs from the total mass of the component), $\alpha$ is the power-law index controlling the behaviour of the inner stellar halo, while $\beta$ controls the behaviour of the outer stellar halo. $J_{0}$ is the scale action characterising the break between the inner and outer stellar halo. The $g$ and $h$ coefficients are linear combinations of the actions. They have the overall effect of controlling the flattening and anisotropy of the density and velocity ellipsoids; $h({\textbf{J}})$ controls the inner stellar halo, while $g(\textbf{J})$ the outer stellar halo. Finally, $\chi$ and $J_{\phi,0}$ control the contribution of rotation to the stellar halo component.

\subsection{Bayesian fitting of the stellar halo model}
We are interested in constraining the parameters of the dark matter contribution to the gravitational potential, and the parameters of the stellar halo DF. The parameters defining the model are thus $\vb*{M}=(\alpha, \beta, J_{0},h_{r}, g_{r}, h_{z}, g_{z}, \chi, \rho_{0}, R_{s}, q)$. We will fit the model using a Bayesian approach, and therefore need to define a likelihood function that returns the probability of the input data $\mathcal{D}$, given the model parameters. We define the individual likelihood $\ell_{i}$ of each star $i$ at its observed phase-space coordinates as the value of the DF at that point, normalized by the total mass of the stellar halo component in the specified model ($\mathcal{N}$) :
\begin{equation}
    \ell_{i}(\vb*{J}) = \frac{f(\vb*{J})}{\mathcal{N}}
\end{equation}
 Thus the total likelihood of the model, $\mathcal{L} = P(\mathcal{D}|\vb*{M})$, is:
\begin{equation}
\label{eq:likelihood}
    \mathcal{L} = \prod_{i} \ell_{i} \Rightarrow \log\mathcal{L} = \sum_{i}\log\ell i = \sum_{i}\log\frac{f(\vb*{J}_{i})}{\mathcal{N}},
\end{equation}
where the sum is over all the stars in the dataset.

According to Bayes' law, the posterior probability is given by:

\begin{equation}
    P(\vb*{M}|\mathcal{D}) = \frac{P(\mathcal{D}|\vb*{M})P(\vb*{M})}{P(\mathcal{D})},
\end{equation}
where $P(\mathcal{D}|\vb*{M})$ is the likelihood (i.e., $\mathcal{L}$ from equation \ref{eq:likelihood}), $P(\vb*{M})$ is the prior, and $P(\mathcal{D})$ is the evidence.
The parameters we fitted for and their prior conditions are stated below. If the conditions are satisfied the prior is 1, otherwise it is 0.
\begin{itemize}
    \item $0<\alpha\leq 3$
    \item $\beta\geq 3$
    \item $J_{0}>0$
    \item $h_{r}>0$, $h_{z}>0$ and $3-h_{r}-h_{z}>0$
    \item $g_{r}>0$,  $g_{z}>0$ and $3-g_{r}-g_{z}>0$
    \item $-1\leq \chi \leq 1$
    \item $\rho_{0}>0$
    \item $R_{s}>0$ 
    \item $0\leq q \leq 1$
\end{itemize}

\noindent The evidence integral gives the probability of the data and is usually difficult to compute:
\begin{equation}
    P(\mathcal{D}) = \int P(\mathcal{D},\vb*{M}) \mathrm{d}\vb*{M}. 
\end{equation}
Given a survey, it would return the probability of all stars being part of that survey. In our case, since all the data points are part of the data set, we can leave out the integral as it will only shift up or down the log-posterior distribution by the same value.

\par The log-posterior distribution is explored using Markov Chain Monte Carlo (MCMC) sampling via the \texttt{emcee} Python package, which uses the MCMC-Hammer algorithm \citep{Foreman-Mackey+2013_MCMCHammer}. MCMC generates samples from the posterior distribution we are interested in. The method relies on constructing a Markov chain for which each subsequent state in the chain is determined through the probability of transition from the previous step to the current state. Our desired posterior probability distribution is then the stationary distribution of the Markov chain. If this stationary distribution exists, it should be achieved after running the MCMC sampler for a large enough number of steps. The samples generated before the convergence should be ignored (this is called the burn-in) and the rest of the samples can then be considered as representative of the stationary distribution.


\section{Test on mock data}  \label{sec:mock_data}
As a first test, we apply our method to a mock stellar halo data set. We describe the construction of the mock data set in \ref{subsec:mock_model}. In the following subsections, we discuss the application of our model to this mock data set.

\subsection{The mock model} \label{subsec:mock_model}
At the core of the mock model lies the total gravitational potential of the mock galaxy and a DF for the stellar halo, as described in Section \ref{sec:method}. The parameters specifying the mock galaxy's gravitational potential and stellar halo DF can be found in Table \ref{tab:mock_bestfit_data_param}. The parameter values of the densities are based on the \citet{Tamm+2012} mass model of M31, with some modification (e.g., we included a scale height for the disk and used slightly different functional forms for the potentials). For the DF, we have chosen MW-like values from \cite{Das+Binney2016_EDF1} and \cite{Das+2016_EDF2}.

\begin{figure}
\centering
\begin{subfigure}{0.38\textwidth}
    \includegraphics[width=.85\linewidth]{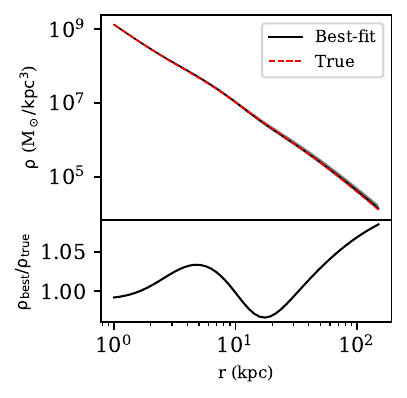}
    \caption{Total density of the best-fit model and true mock galaxy.}
    \label{fig:mock_totdens}
\end{subfigure}
\begin{subfigure}{0.38\textwidth}
    \includegraphics[width=.85\linewidth]{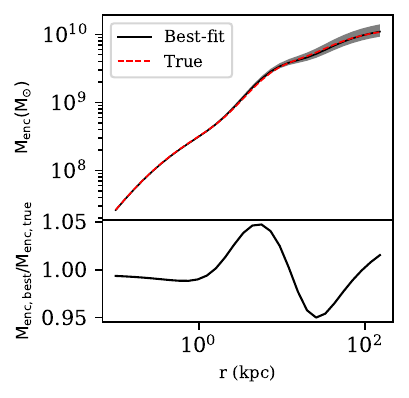}  
    \caption{Total enclosed mass profiles for the best-fit model and true mock galaxy.}
    \label{fig:mock_encmass}
\end{subfigure}
\caption{Comparison between the true (red line) and best-fit (dotted, black line) profiles for the mock model (a) density (b) total enclosed mass profile. The $1\sigma$ confidence intervals are shown in grey. The best-fit model provides a very good fit to the true mock galaxy.}
\label{fig:mock_dens_encmass_tot}
\end{figure}

\begin{figure}
\centering
\begin{subfigure}{0.38\textwidth}
    \includegraphics[width=.85\linewidth]{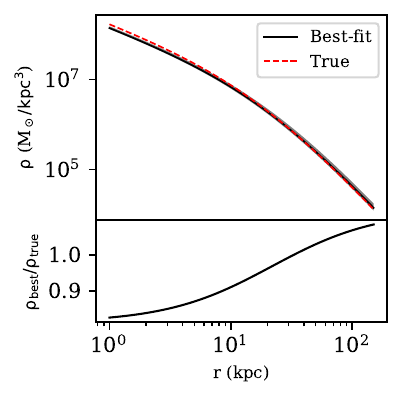}  
    \caption{DM density of the best-fit model and true mock galaxy.}
    \label{fig:mock_DMdens}
\end{subfigure}

\begin{subfigure}{0.38\textwidth}
    \includegraphics[width=.85\linewidth]{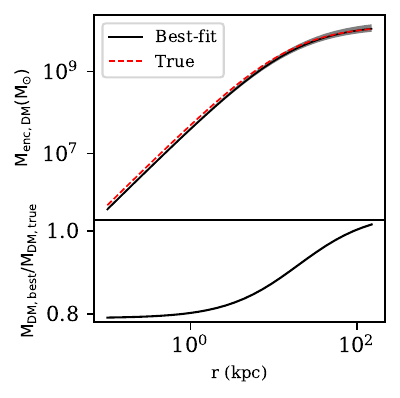}  
    \caption{DM enclosed mass profiles for the best-fit model and true mock galaxy.}
    \label{fig:mock_DMencmass}
\end{subfigure}

\caption{Comparison between the true (red line) and best-fit (dotted, black line) profiles for the mock model (a) DM density (b) DM total enclosed mass profile. The $1\sigma$ confidence intervals are shown in grey. The best-fit model provides a very good fit to the true mock galaxy.}
\label{fig:mock_dens_encmass_DM}
\end{figure}

The resulting total density profile of the mock galaxy can be seen in Figure \ref{fig:mock_totdens} (the red, dashed line). This corresponds to the total gravitational potential arising from all the galactic components (bulge, disk, and DM halo). The overall profile is described by an (approximately) single power law. The red, dashed line in Figure \ref{fig:mock_DMdens} shows the DM halo density profile of the mock model, which follows an NFW profile \cite{Navarro+1997_NFW}.

We sampled $n_{\mathrm{stars}} = 15000$ stars from the galaxy model to create the mock data set ($\vb*{x},\vb*{v}$). In future work, we plan to apply this method to real M31 data, and therefore sample a number of stars comparable to the number of stellar halo samples (e.g. observed with the PAndAS survey). The sampling is done in \texttt{AGAMA} through an adaptive rejection sampling method, which generates $N_{\mathrm{stars}}$ $(\vb*{x_{i},v_{i}})$ such that their density at a point $\vb*{x}$ is proportional to the value of the DF at that point \citep[see the extended \texttt{AGAMA} documentation,][]{Vasiliev2018_Agamaext} for the detailed sampling procedure description.

\begin{table*}
  \caption{\label{tab:mock_bestfit_data_param}In the third column, we show the parameters of the gravitational potential and distribution function used to create the mock data sets. The parameter values of the bulge, disk, and DM halo potentials are based on the \protect\citet{Tamm+2012} M31 mass model (with some modifications). The parameters for the distribution function are taken to be Milky Way-like values from \protect\cite{Das+Binney2016_EDF1} and \protect\cite{Das+2016_EDF2}. The last column shows the best-fit values (and associated 68\% confidence intervals) of the recovered parameters by our dynamical model. An asterisk `*' indicates that the parameter has been fixed prior to the run.}
  \centering 
  \begin{threeparttable}
    \begin{tabular*}{0.8\linewidth}{@{\extracolsep{\fill}}cccc}
\midrule
    Component  & Parameter & Value mock & Value best-fit\\
     \midrule
DM halo potential  &   
$\begin{aligned}
    \rho_{0,\mathrm{DM}}\\
    R_{s}
\end{aligned}
$        &  
$\begin{aligned}
    & 1.1E+7 \textrm{  }\mathrm{M_{\odot}}/\mathrm{kpc^{3}} \\
    & 17 \textrm{ kpc}
\end{aligned}$ 

 &  
$\begin{aligned}
    & {0.73 \mathrm{E}+7}^{+9.65\mathrm{E}+5}_{-9.01\mathrm{E}+5} \textrm{  } \mathrm{M}_{\odot}/\mathrm{kpc^{3}} \\
    & 20.3^{+1.38}_{-1.22} \textrm{ kpc}
\end{aligned}$ 

\\
    \cmidrule(l  r ){1-4}
    Bulge potential & $ \begin{aligned}
    M_{\mathrm{bulge}} \\
    R_{b} \\
    n_{b} \\
    q_{b} \\
    \end{aligned} $ &
    $\begin{aligned} 
    & 3.1 \mathrm{E}+10 \: \mathrm{M}_{\odot} \\
    & 1.155 \textrm{ kpc}\\
    & 2.7 \\
    & 0.72 \\
    \end{aligned}$  

     &
    $\begin{aligned} 
    & 3.09\mathrm{E}+10^{+4.87\mathrm{E}+8}_{-4.66\mathrm{E}+8} \: \mathrm{M}_{\odot} \\
    & 1.155* \textrm{ kpc}\\
    & 2.7* \\
    & 0.72* \\
    \end{aligned}$  
    
\\
    \cmidrule(l  r ){1-4}
    disk potential & $ \begin{aligned}
    M_{\mathrm{disk}} \\
    R_{d} \\
    h \\
    n_{d} \\
    \end{aligned} $ &
    $\begin{aligned} 
    & 5.6 \mathrm{E}+10 \: \mathrm{M}_{\odot} \\
    & 2.57 \textrm{ kpc}\\
    & 0.4 \textrm{ kpc} \\
    & 1.2 \\
    \end{aligned}$ 
     &
    $\begin{aligned} 
    & {7.03\mathrm{E}+10}^{+3.28\mathrm{E}+9}_{-3.22\mathrm{E}+9} \: \mathrm{M}_{\odot} \\
    & 2.57* \textrm{ kpc}\\
    & 0.4* \textrm{ kpc} \\
    & 1.2* \\
    \end{aligned}$ 

\\
    \cmidrule(l  r ){1-4}
     Stellar halo DF & $ \begin{aligned}
    M_{0} \\
    \alpha \\
    \beta \\
    J_{0} \\
    h_{r} \\
    h_{z} \\
    g_{r} \\
    g_{z} \\
    \chi \\
    J_{0,\phi}
    \end{aligned} $ &
    $\begin{aligned} 
    & 1\\
    & 2.5 \\
    & 5.5 \\
    & 8000 \: \mathrm{kpc\,km\,s^{-1}} \\
    & 0.75 \\
    & 1.7 \\
    & 0.88 \\
    & 1.1 \\
    & 0.5 \\
    & 1 \\
    
    \end{aligned}$  
    &
    $\begin{aligned} 
    & 1*\\
    & 2.5^{+0.01}_{-0.01} \\
    & 5.63^{+0.24}_{-0.21} \\
    & 8440.93^{+998.68}_{-854.43} \: \mathrm{kpc\,km\,s^{-1}} \\
    & 0.76^{+0.01}_{-0.01} \\
    & 1.68^{+0.01}_{-0.1} \\
    & 0.87^{+0.02}_{-0.02} \\
    & 1.09^{+0.03}_{-0.03} \\
    & 0.5^{+0.01}_{-0.01} \\
    & 1* \\
    
    \end{aligned}$ 
\\ 
    \midrule
    \end{tabular*}
\end{threeparttable}

 \end{table*}

\begin{figure*}
    \centering
    \includegraphics[width=\textwidth]{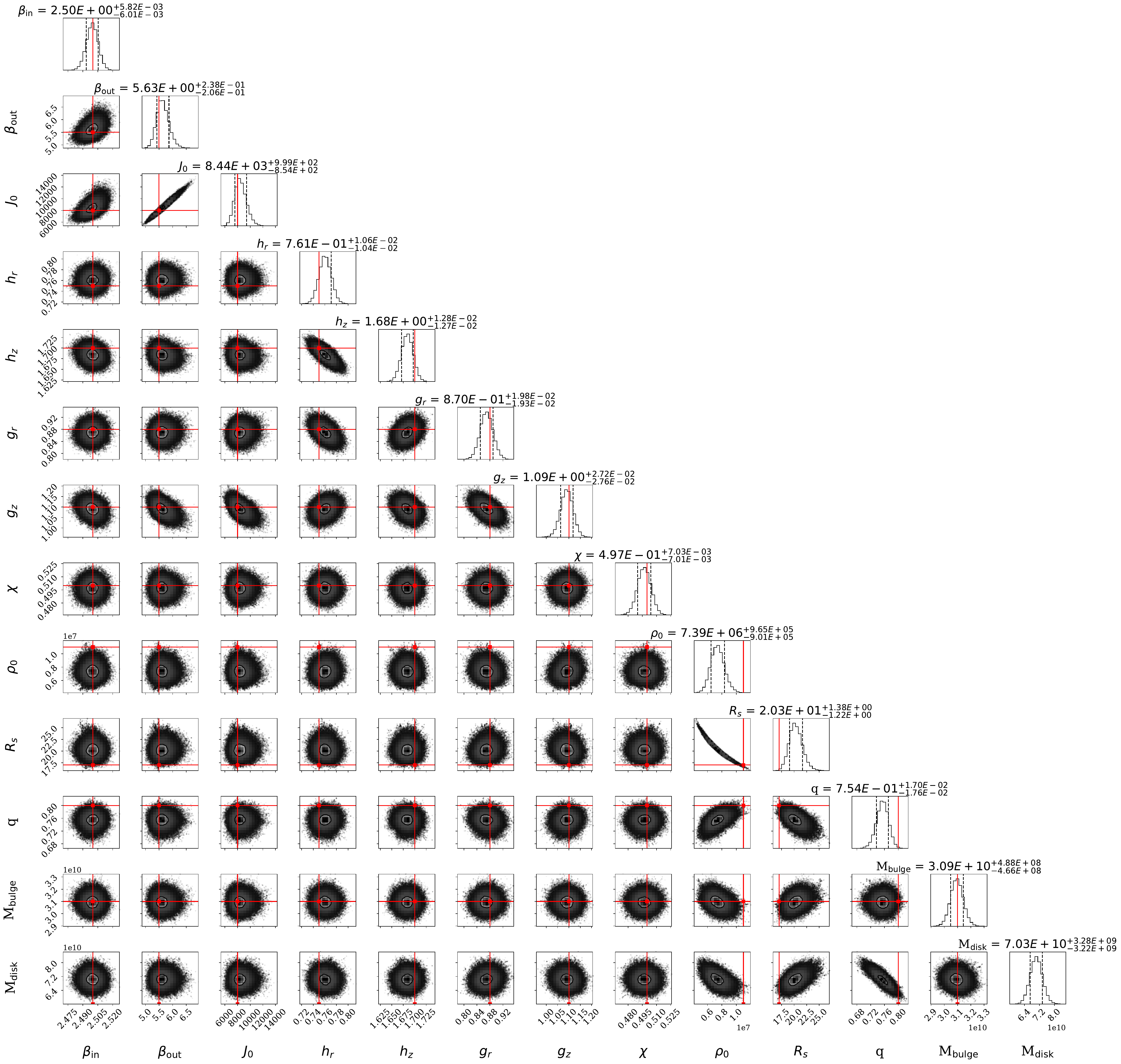}
    \caption{Corner plot for the $\texttt{emcee}$ chains generated when fitting the mock galaxy sample. The red lines represent the true values from Table \ref{tab:mock_bestfit_data_param}. The titles of each histogram show the recovered value for the parameters and the associated 1$\sigma$ uncertainties. The diagonal plots show 1D histograms of each parameter, marginalized over all other parameters. The off-diagonal show 2D histograms demonstrating correlations for each combination of two parameters, marginalized over the remaining parameters.}
    \label{fig:corner_mock_15000}
\end{figure*}


\vspace{12pt}

Next, we run the fitting procedure with the mock data set. The parameters specifying the model (i.e., the parameters we want to recover) are ($\alpha, \beta,J_{0},h_{r},h_{z},g_{r},g_{z},\chi,\rho_{0},R_{s},q,M_{\mathrm{bulge}},M_{\mathrm{disk}}$). As we are focusing on understanding the DM halo and stellar halo DF, we have chosen two free parameters only ($M_{\mathrm{bulge}}$ and $M_{\mathrm{disk}}$) for the inner gravitational potential.

\par We ran $\texttt{emcee}$ for 30,000 steps with 26 walkers for a mock data set on 16 cores, which took $\approx 36$ hours to run. The number of walkers chosen is the minimum number required to ensure their linear independence, which is suggested to be twice the number of the parameters of the model. To visualise the results of the MCMC analysis, we use the \texttt{Python} package \texttt{corner.py}. The diagonal plots show the distributions of each of the model parameters marginalised over all other parameters, while the off-diagonal plots show joint distributions of pairs of parameters, marginalised over the remaining parameters. Figure \ref{fig:corner_mock_15000} shows the corner plot of the samples generated in our $\texttt{emcee}$ chains. A summary of the best-fit parameters and associated confidence intervals can be found in Table \ref{tab:mock_bestfit_data_param}. As expected, the best-fit parameters lie within the $1\sigma$ confidence intervals 65\% of the time and, within $2\sigma$, 95\% of the time. Furthermore, discrepancies between the true and best-fit values come also from the fact that we are investigating a random sample of the true underlying population. Therefore, these recovered parameters are representative of the sample, not of the population. Nonetheless, results show a very good overall fit. 
 
\par The confidence intervals are very narrow. The sample size we use is (relatively) large which reduces the variance in the estimates of the model's parameters and can result in narrower confidence intervals. Furthermore, the use of informative priors can have the effect of significantly narrowing the uncertainty in the posterior distribution, leading to narrower confidence intervals. Moreover, the MCMC algorithm has converged quickly ($\approx 500$ steps), which can also have the effect of narrowing the confidence intervals by providing a more accurate posterior determination. Last but not least, low noise on data improves the preciseness of the parameter estimates, therefore leading to narrower confidence intervals (our mock data has no associated error or noise).

Correlations can also be seen between some parameters. $J_{0}$ positively correlates with $\beta$, picking out solutions with the same number of stars at some outer radius. There is also a strong negative correlation between the DM parameters $R_{s}$ and $\rho_{0}$, similarly because this conserves the amount of total DM matter mass at some outer radius.

\subsection{Mass distribution} 
The density profile (Figure \ref{fig:mock_totdens}) recovered by the best-fit model (dashed, black line, corresponding to the total mass distribution), shows very good agreement with the true total density (red line), with the confidence intervals being very narrow. The recovered DM density profile can be seen in Figure \ref{fig:mock_DMdens} as the black line and shows good agreement with the true DM density profile (red line).

Figures \ref{fig:mock_encmass} and \ref{fig:mock_DMencmass} show the spherically averaged total enclosed mass and dark matter mass profiles of the true and best-fit models. Overall, there is a very good agreement between the true model and the recovered best-fit model, with values agreeing within the confidence intervals. The confidence intervals for the total enclosed mass profiles are wider for radii $r\gtrapprox30$ kpc. On the other hand, for the DM enclosed mass, the confidence intervals are wider at lower radii $r\lessapprox10$ kpc.


\subsection{The stellar halo distribution function}


\begin{figure}
    \includegraphics[width=0.35\textwidth]{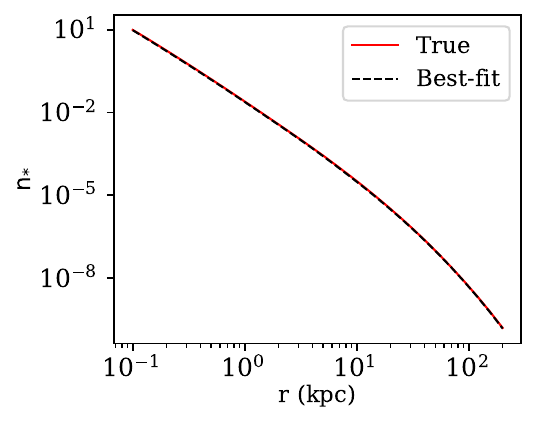}
    \caption{Comparison between the true (red line) and best-fit (dotted, black line) profiles for the mock model's stellar halo density (with total mass normalized to unity) profile. The two show very good agreement.}
    \label{fig:mock_shalo_ndens}
\end{figure}

\subsubsection{Number density}
Figure \ref{fig:mock_shalo_ndens} shows the stellar halo density profile for the best-fit model and for the true mock galaxy. The density corresponds to the zeroth moment of the DF:
\begin{equation}
    n(\vb*{x}) = \iiint \mathrm{d^{3}v}f\big(\vb*{J}(\vb*{x},\vb*{v})\big).
\end{equation}
Since we are not doing self-consistent modelling, the total mass of the stellar halo is unknown. Therefore, we set the DF mass normalization, from which the total mass of the component is computed, to unity when modelling the stellar halo. Changing the total mass value does not change the plots qualitatively, it only acts as a rescaling parameter.

It can be seen that both the true and best-fit density profiles follow a power law distribution, with the inner slope slightly less steep than the outer one. Both profiles agree very well.  

\subsubsection{Velocity distribution}
\begin{figure}
\centering
    \includegraphics[width=0.35\textwidth]{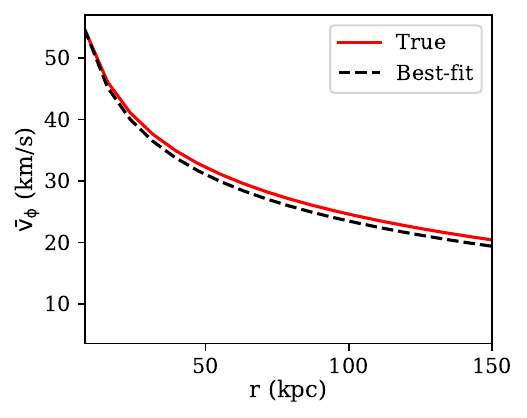}
    \caption{Comparison of average $v_{\phi}$ (in cylindrical coordinates) vs radius. It can be seen that the stellar halo has an overall net rotation.}
    \label{fig:mock_vphi_shalo}
\end{figure}

Figure \ref{fig:mock_vphi_shalo} shows the radial $\mathrm{v_{\phi}}$ velocity profile for both the best-fit and true models. The fact that this component is non-zero indicates an overall net rotation of the stellar halo. This agrees with the non-zero rotation fraction $\chi$ in the DF (see table \ref{tab:mock_bestfit_data_param}).

\begin{figure}
    \includegraphics[width=0.35\textwidth]{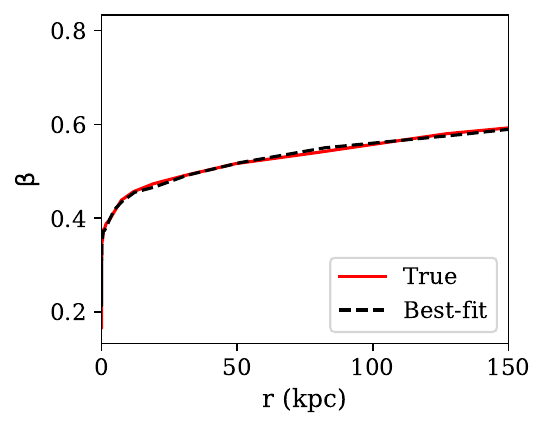}
    \caption{The spherical anisotropy parameter $\beta$ against radius.}
    \label{fig:mock_anisotropy_shalo}
\end{figure}
Next, we evaluate the velocity anisotropy ($\mathrm{\beta}$) profile. In a spherical, galactocentric potential, the form of the $\mathrm{\beta}$ parameter is, as defined by \citet{Binney&Tremaine1987_GalDynamics},

\begin{equation}
    \beta = 1-(\sigma_{\theta}^{2}+\sigma_{\phi}^2)/\sigma_{r}^{2},
\end{equation}
where $\sigma_{\theta}$, $\sigma_{\phi}$, and $\sigma_{r}$ are the velocity dispersions in a spherical coordinate system.

This parameter quantifies the degree of radial ($\mathrm{\beta}>0$) or tangential ($\mathrm{\beta}<0$) bias of stellar orbits, while $\mathrm{\beta} = 0$ indicates isotropy. The profiles were computed by generating ($\vb*{x},\vb*{v}$) samples from both the mock data and best-fit models. The data has been binned in radial bins of a given width $\Delta r$, and the dispersions of each velocity component ($v_{r},v_{\theta},v_{\phi}$) have been computed in the corresponding bins.

Figure \ref{fig:mock_anisotropy_shalo} show the anisotropy profiles of both true and best-fit models, which are in good agreement. It can be seen that $\beta$ increases rapidly from $\sim 0$ to $\sim 0.4$ within a few kpc, and then approximately plateaus to $0.6$ in the outer halo. The stellar halo is therefore moderately radial anisotropic throughout.

\subsubsection{Stellar halo flattening} \label{subsec:mock_shalo_flattening}

\begin{figure}
    \includegraphics[width=0.35\textwidth]{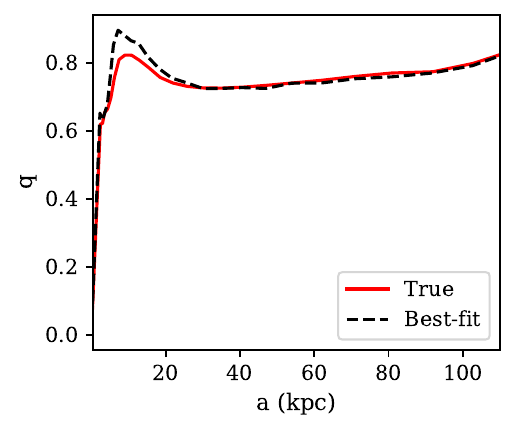}  
    \caption{Comparison between the true (red line) and best-fit (dotted, black line) profiles for the mock model's stellar halo flattening vs. elliptical radius.}
    \label{fig:mock_shalo_flattening}
\end{figure}

\par Figure \ref{fig:mock_shalo_flattening} compares the axis ratios of the stellar haloes. To compute the axis ratio $q$, we fit ellipses to stellar halo density contours. The resulting axis ratio at a given elliptical radius, $a = \sqrt{x^2+y^2+(z/q)^2}$, is defined by the ratio between the minor and major axis of the fitted ellipse. It can be seen that for both the true and the best-fit stellar halo, the axis ratio increases steeply until $a\approx 10$ kpc, after which it approximately plateaus at $q\approx 0.75$. This likely reflects the impact of the disk on the stellar halo shape in the central region and the impact of the dark matter halo further out.


\section{Application to Auriga haloes} \label{sec:auriga}

In this section, we apply our stellar halo model to the Auriga simulations. We first discuss the simulations and then present the fits to three of the Auriga haloes.

\subsection{The Auriga simulations}
The Auriga simulations are a suite of 30 high-resolution magneto-hydrodynamical zoom-in cosmological simulations in a $\Lambda$CDM Universe. The parent haloes are selected from the dark matter-only EAGLE simulation \citep{Schaye+2015_EAGLE} based on an isolation criterion at $z=0$. The selected haloes are then re-simulated at a higher resolution through the zoom-in technique with the moving mesh code AREPO given an extensive galaxy formation model, which includes processes such as gas cooling and heating, star formation, supernova feedback, and black hole growth.

\begin{figure*}
    \centering
    \includegraphics[width=0.9\textwidth]{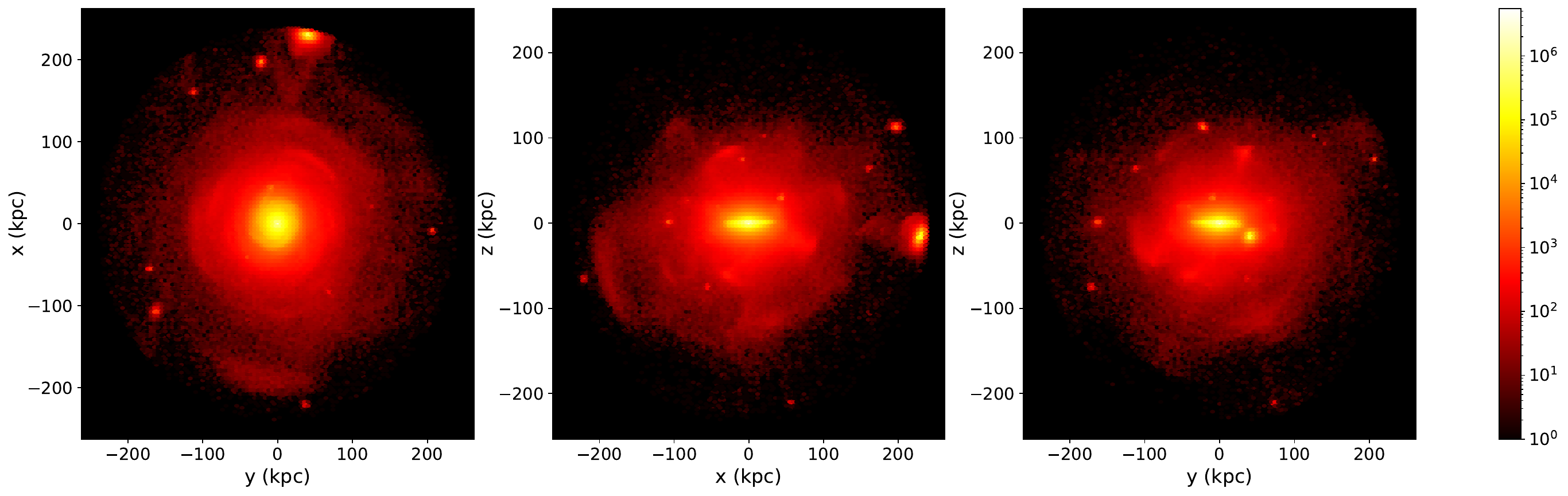}
    \caption{Hexbin projections of halo 24 of the Auriga simulations in the $x$-$y$, $x$-$z$, and $y$-$z$ planes. The plots show the stellar particles only (in-situ, ex-situ, and part of subhaloes).}
    \label{fig:halo21}
\end{figure*}

Six Auriga haloes have been processed through the Auriga2PAndAS pipeline \citep{Thomas+2021_Auriga2Pandas} to create PAndAS-like mocks. In this paper, we investigate three of these haloes (haloes 21, 23, and 24). These haloes have numerous substructures present and have also undergone a recent major merger. Details of the parameters of the investigated haloes can be found in table \ref{tab:Auriga_haloes_param}. While all the haloes chosen are M31-like, we have chosen them to reflect different degrees of "busyness", i.e., number of streams, substructure, departure from axisymmetry, etc. Figure \ref{fig:halo21} shows, for illustration purposes, a density map projected into the $x$-$y$, $x$-$z$, and $y$-$z$ planes for halo 24. The stellar content is split into in-situ, accreted, or as part of sub-haloes. In this work, we are defining the stellar halo as the accreted stellar content.
\par Next, we applied our fitting method to each of the three Auriga haloes. The sample size for each halo is $n_{\mathrm{stars}}=15000$. Table \ref{tab:Auriga_results} shows the results with the best-fit parameters. Parameters that were fixed prior to the runs are marked by an asterix *. In the following subsections, we present the results in more detail.

\begin{table}
  \caption{\label{tab:Auriga_haloes_param}Table of parameters of the investigated Auriga haloes. The columns are: 1) halo name, 2) virial mass, 3) virial radius, 4) stellar mass, 5) number of significant progenitors which have contributed 90 $\%$ of the accreted stellar mass of the stellar halo. The parameters have been taken from Table 1 of \citet{Monachesi+2019}.}
 \centering 
  \begin{threeparttable}
    \begin{tabular*}{\linewidth}{@{\extracolsep{\fill}}ccccc}
\midrule
Halo  & $\frac{M_{\mathrm{vir}}}{(10^{10}\mathrm{M}_{\odot})}$ & $\frac{R_{\mathrm{vir}}}{(\mathrm{kpc})}$ & $\frac{M_{\mathrm{*}}}{(10^{10}\mathrm{M}_{\odot})}$ & $N_{\mathrm{sp}}$  \\
     \midrule
$\begin{aligned}
    & \mathrm{Au21} \\
    & \mathrm{Au23} \\
    & \mathrm{Au24} 
\end{aligned}
$        &  
$\begin{aligned}
    & 145.09 \\
    & 157.53  \\
    & 149.17 
\end{aligned}$ 
        &  
$\begin{aligned}
    & 238.64 \\
    & 245.27  \\
    & 240.85 
\end{aligned}$
        &  
$\begin{aligned}
    & 8.65 \\
    & 9.80  \\
    & 7.66  
\end{aligned}$
        &  
$\begin{aligned}
    & 4 \\
    & 8  \\
    & 8  
\end{aligned}$
\\ 
    \midrule
    \end{tabular*}
\end{threeparttable}
 \end{table}

\begin{table*}
  \caption{\label{tab:Auriga_results}The median and 68$\%$ confidence intervals of the recovered parameters for the three Auriga stellar haloes investigated. Parameters marked with an '*' have been fixed prior to the runs.}
  \begin{threeparttable}
    \begin{tabular*}{\linewidth}{@{\extracolsep{\fill}}cccc}
\midrule
Parameter  & Auriga 21 & Auriga 23 & Auriga 24 \\
     \midrule
$\begin{aligned}
    & R_{b}\\
    & n_{b}\\
    & q_{b}\\
    & R_{d}\\
    & h\\
    & n_{d}\\
    & M_{0} \\
    & J_{0,\phi} \\
    & \alpha \\
    & \beta \\
    & J_{0} \\
    & h_{r} \\
    & h_{z} \\
    & g_{r} \\
    & g_{z} \\
    & \chi \\
    & \rho_{0}\\
    & R_{s} \\
    & q\\
    & \mathrm{M_{bulge}}\\
    & \mathrm{M_{disk}}
\end{aligned}
$        &  
$\begin{aligned}
    & 1.155^{*} \textrm{ kpc} \\
    & 2.7^{*} \\
    & 0.72^{*} \\
    & 2.57^{*} \textrm{ kpc} \\
    & 0.4^{*} \textrm{ kpc} \\
    & 1.2^{*} \\
    & 1^{*} \\
    & 1^{*} \\
    & 0.65^{+0.08}_{-0.08} \\
    & 3.00^{+0.01}_{-0.01}  \\
    & 375.74^{+25.34}_{-23.50} \: \mathrm{kpc\,km\,s^{-1}} \\
    & 0.37^{+0.06}_{-0.06} \\
    & 2.46^{+0.08}_{-0.08}\\
    & 0.91^{+0.01}_{-0.01}\\
    & 1.19^{+0.01}_{-0.01}\\
    & -0.50^{+0.01}_{-0.01}\\
    & {5.77\mathrm{E}+6}^{+8.31\mathrm{E}+5}_{-7.57\mathrm{E}+5} \textrm{  } \mathrm{M}_{\odot}/\mathrm{kpc^{3}} \\
    & 23.56^{+1.36}_{-1.20} \textrm{ kpc}\\
    & 0.93^{+0.01}_{-0.02} \\
    & {1.15\mathrm{E}+10}^{+1.06\mathrm{E}+9}_{-1.03\mathrm{E}+9} \: \mathrm{M}_{\odot}\\
    & {6.46\mathrm{E}+10}^{+3.64+\mathrm{E}+9}_{-3.59\mathrm{E}+9} \: \mathrm{M}_{\odot}
\end{aligned}$ 

        &  
$\begin{aligned}
    & 1.155^{*} \textrm{ kpc} \\
    & 2.7^{*} \\
    & 0.72^{*} \\
    & 2.57^{*} \textrm{ kpc} \\
    & 0.4^{*} \textrm{ kpc} \\
    & 1.2^{*} \\
    & 1^{*} \\
    & 1^{*} \\
    & 1.96^{+0.02}_{-0.004} \\
    & {1.41\mathrm{E}+15}^{+1.48\mathrm{E}+13}_{-1.5\mathrm{E}+13}  \\
    & {5.14\mathrm{E}+19}^{+2.70\mathrm{E}+18}_{-2.4\mathrm{E}+18} \: \mathrm{kpc\,km\,s^{-1}} \\
    & 0.78^{+0.001}_{- 0.03} \\
    & 2.05^{+0.02}_{-0.006}\\
    & 0.72^{+0.08}_{-0.03}\\
    & 0.87^{+0.01}_{-0.07}\\
    & -0.61^{+0.01}_{-0.001}\\
    & {2.12\mathrm{E}+7}^{+2.49\mathrm{E}+6}_{-4.11\mathrm{E}+6} \textrm{  } \mathrm{M}_{\odot}/\mathrm{kpc^{3}} \\
    & 17.20^{+1.20}_{-1.15} \textrm{ kpc}\\
    & 0.58^{+0.03}_{-0.006} \\
    & {2.29\mathrm{E}+10}^{+1.08\mathrm{E}+9}_{-3.19\mathrm{E}+9} \: \mathrm{M}_{\odot}\\
    & {2.52\mathrm{E}+10}^{+1.33\mathrm{E}+9}_{-1.60\mathrm{E}+8} \: \mathrm{M}_{\odot}
\end{aligned}$ 

        &  
$\begin{aligned}
    & 1.155^{*} \textrm{ kpc} \\
    & 2.7^{*} \\
    & 0.72^{*} \\
    & 2.57^{*} \textrm{ kpc} \\
    & 0.4^{*} \textrm{ kpc} \\
    & 1.2^{*} \\
    & 1^{*} \\
    & 1^{*} \\
    & 1.56^{+0.02}_{-0.02} \\
    & 3.22^{+0.04}_{-0.04}  \\
    & 2818.61^{+136.77}_{-126.57} \: \mathrm{kpc\,km\,s^{-1}} \\
    & 1.07^{+0.02}_{-0.02} \\
    & 1.90^{+0.02}_{-0.02}\\
    & 0.77^{+0.01}_{-0.01}\\
    & 1.46^{+0.01}_{-0.01}\\
    & -0.48^{+0.01}_{-0.01}\\
    & {1.45\mathrm{E}+7}^{+1.33\mathrm{E}+6}_{-1.26\mathrm{E}+6} \textrm{  } \mathrm{M}_{\odot}/\mathrm{kpc^{3}} \\
    & 18.24^{+0.73}_{-0.67} \textrm{ kpc}\\
    & 0.72^{+0.01}_{-0.02} \\
    & {2.55\mathrm{E}+10}^{+9.45\mathrm{E}+8}_{-9.32\mathrm{E}+8} \: \mathrm{M}_{\odot}\\
    & {1.01\mathrm{E}+10}^{+2.18\mathrm{E}+9}_{-2.12\mathrm{E}+9} \: \mathrm{M}_{\odot}
\end{aligned}$ 

\\ 
    \midrule
    \end{tabular*}
\end{threeparttable}
 \end{table*}

\subsection{Mass distribution} \label{subsec:mass_distribution_Auriga}
\begin{figure*}
\centering
\begin{subfigure}{0.95\textwidth}
    \includegraphics[width=.95\linewidth]{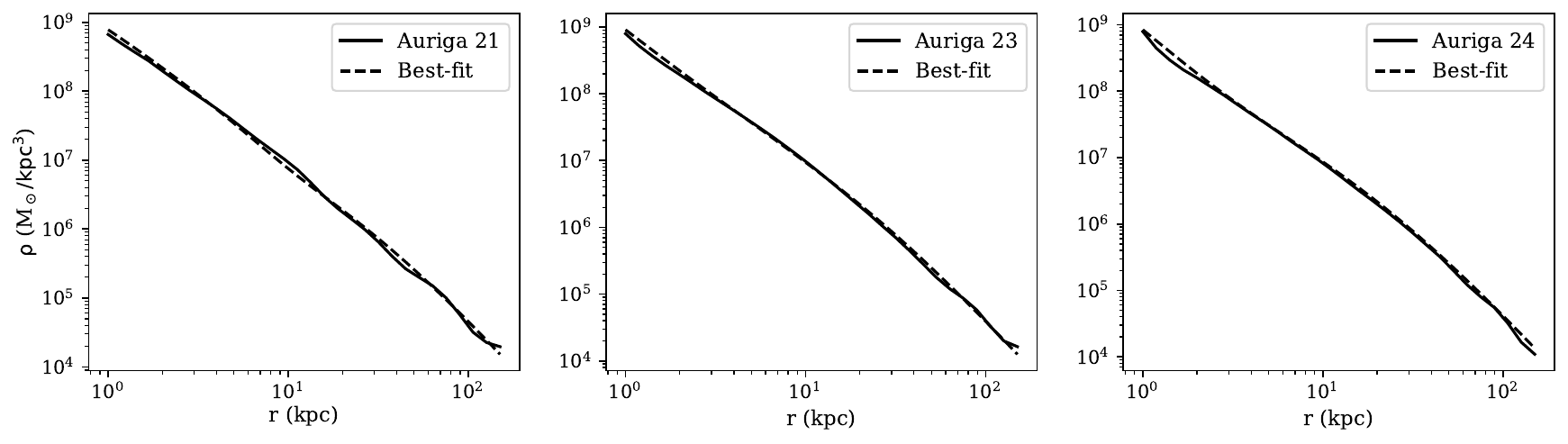}
    \caption{Total density profiles for the best-fit model and true Auriga haloes.}
    \label{fig:totdens_Auriga}
\end{subfigure}
\begin{subfigure}{0.95\textwidth}
    \includegraphics[width=.95\linewidth]{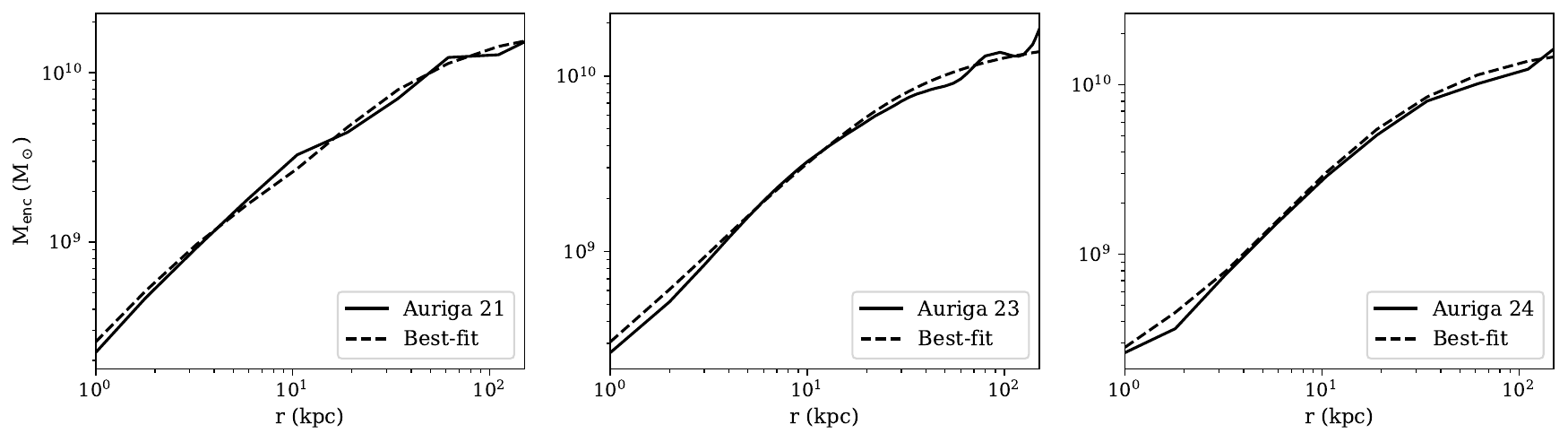}  
    \caption{Total enclosed mass profiles for the best-fit model and true Auriga haloes}
    \label{fig:encmass_tot_Auriga}
\end{subfigure}
\begin{subfigure}{0.95\textwidth}
    \includegraphics[width=.95\linewidth]{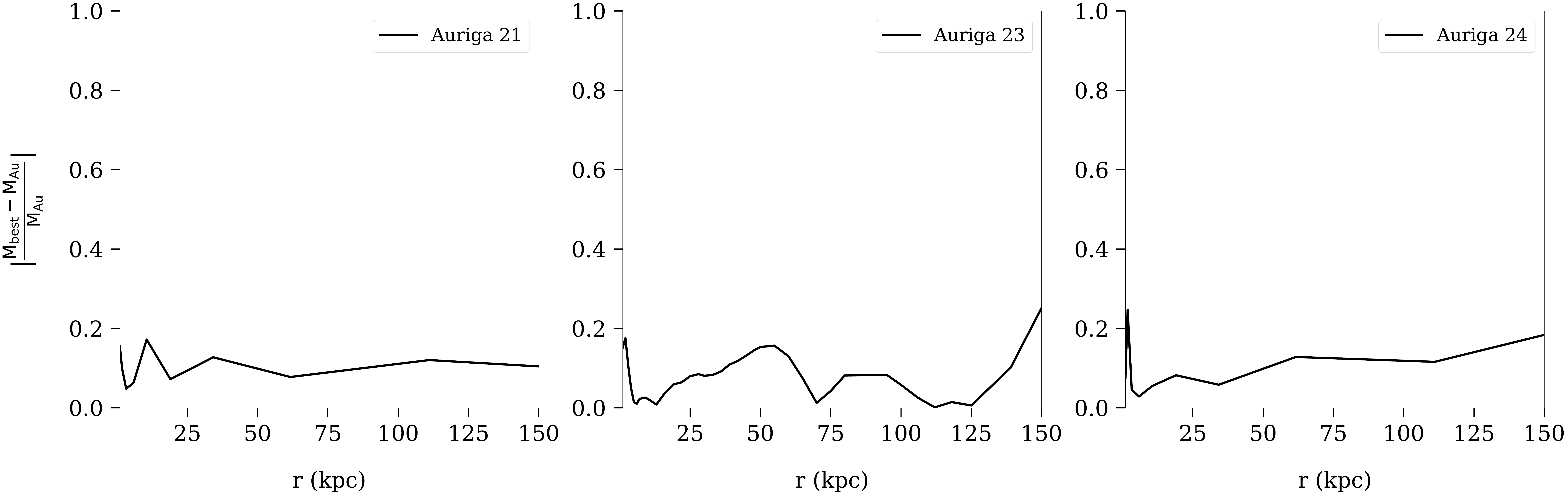}  
    \caption{Fractional difference of the total enclosed mass between the best-fit model and true Auriga haloes.}
    \label{fig:fracdiff_totencmass_Auriga}
\end{subfigure}

\caption{Comparison between the true (continuous, black line) and best-fit (dotted, black line) profiles for the three Auriga haloes investigating (a) total density, (b) total enclosed mass. Subfigure (c) shows the fractional difference between the true and best-fit enclosed mass profiles. There is very good agreement between the best-fit model and the true Auriga halos, with the model incurring a mass error which is always below $20\%$.}
\label{fig:dens_encmass_fracdiff_Auriga}
\end{figure*}

\begin{figure*}
\centering
\begin{subfigure}{0.95\textwidth}
    \includegraphics[width=.95\linewidth]{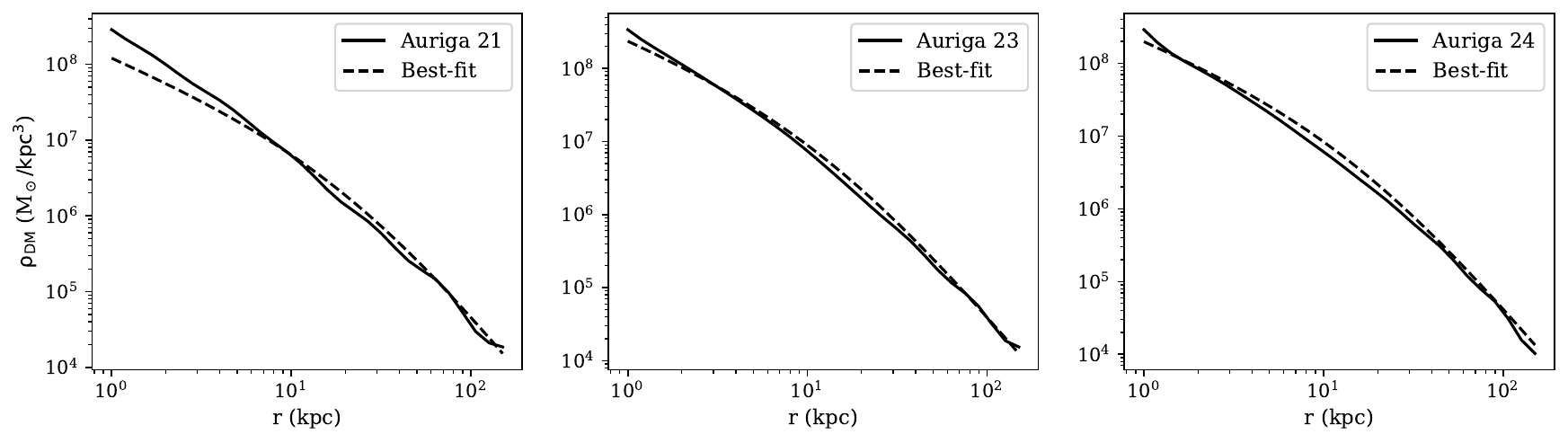}
    \caption{DM density profiles for the best-fit model and true Auriga haloes.}
    \label{fig:dens_DM_Auriga}
\end{subfigure}
\begin{subfigure}{0.95\textwidth}
    \includegraphics[width=.95\linewidth]{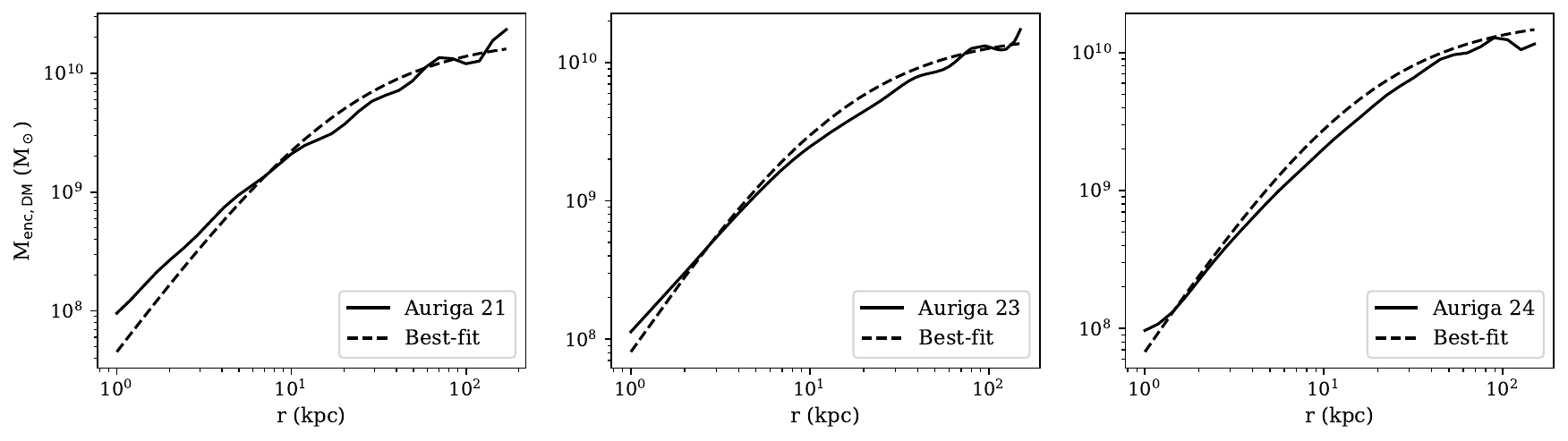}  
    \caption{DM enclosed mass profiles for the best-fit model and true Auriga haloes}
    \label{fig:encmass_DM_Auriga}
\end{subfigure}
\begin{subfigure}{0.95\textwidth}
    \includegraphics[width=.95\linewidth]{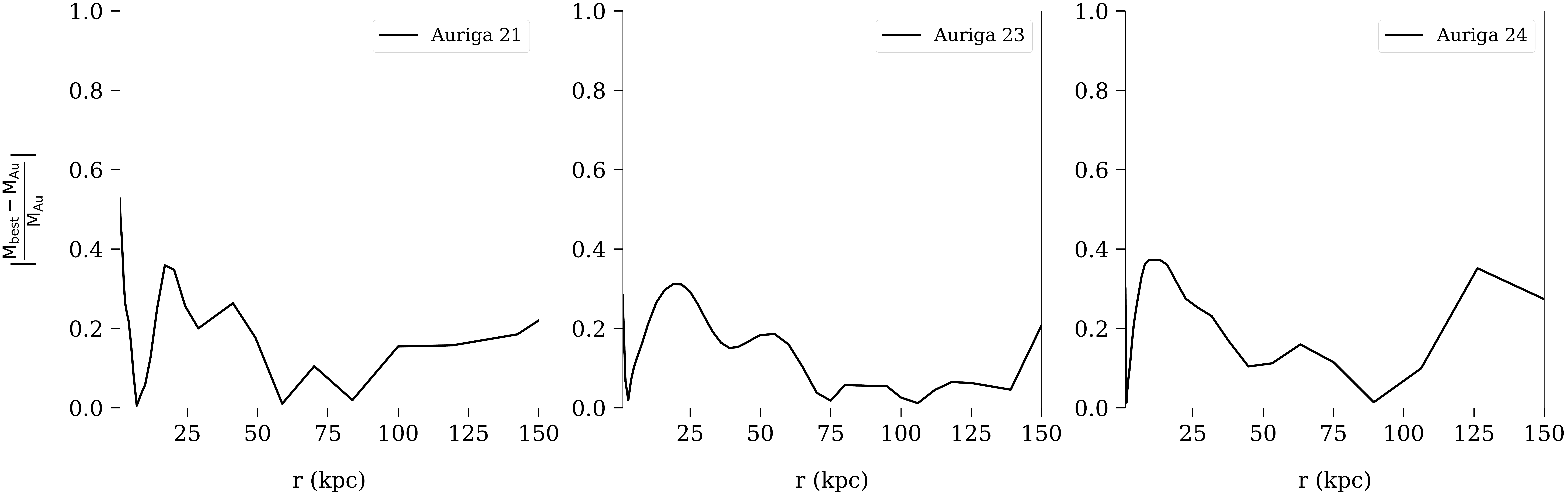}  
    \caption{Fractional difference of the DM enclosed mass between the best-fit model and true Auriga haloes.}
    \label{fig:fracdiff_DMencmass_Auriga}
\end{subfigure}

\caption{Comparison between the true (continuous, black line) and best-fit (dotted, black line) profiles for the three Auriga haloes investigating (a) DM density, (b) DM enclosed mass. Subfigure (c) shows the fractional difference between the true and best-fit DM enclosed mass profiles. The fits agree well, but it can be seen that the best-fit departs more form the true Auriga DM haloes at lower radii.}
\label{fig:dens_encmass_fracdiff_DM_Auriga}
\end{figure*}

\begin{figure*}
    \centering
    \includegraphics[width=\textwidth]{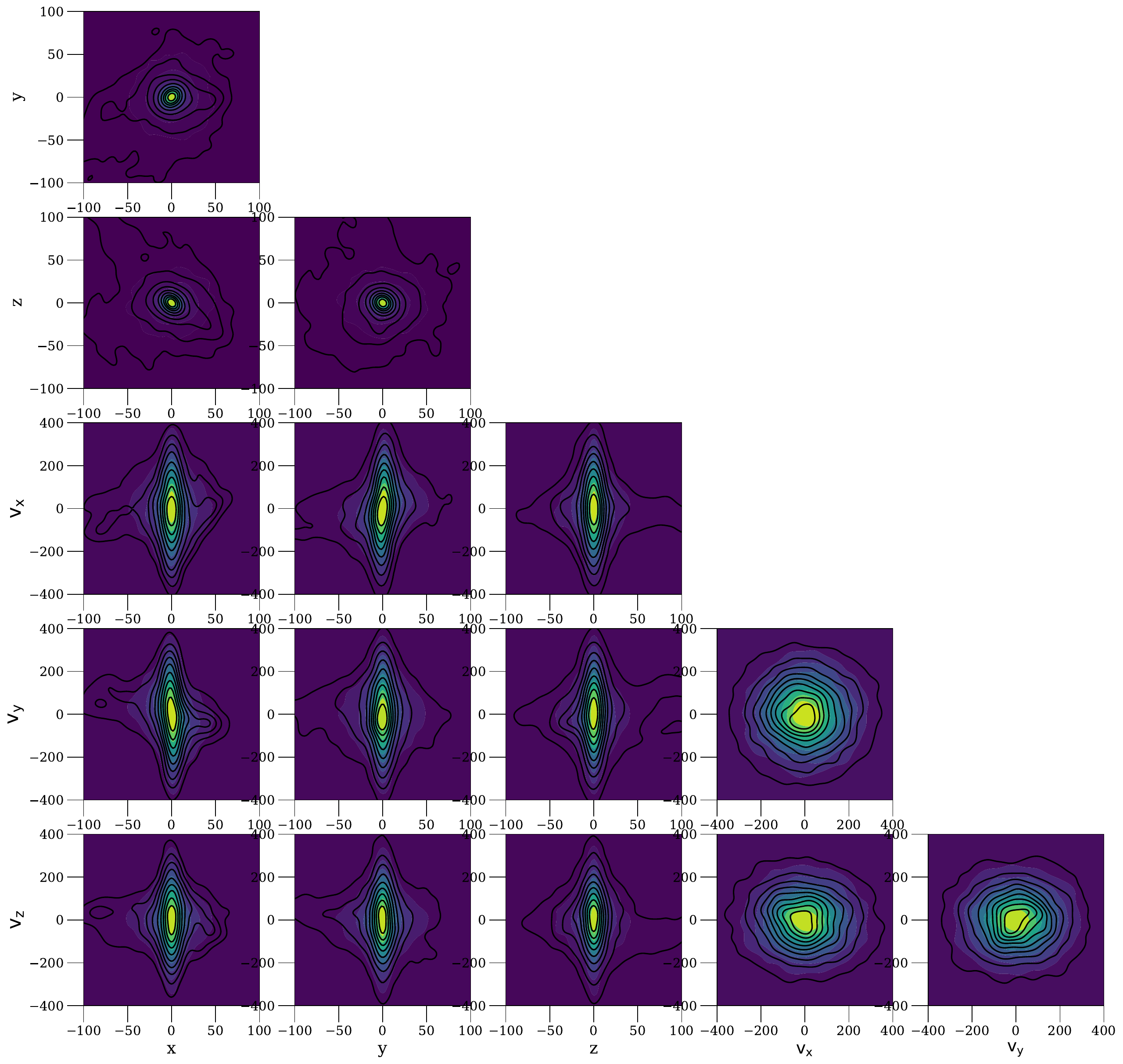}
    \caption{The color-filled contours show the 2D distributions of the ($\mathbf{x},\mathbf{v}$) phase-space of the best-fit model for the Auriga 24 halo. The black contours show the same 2D distributions but for the Auriga 24 stellar halo stars. The contours have been created with a sample of 10,000 stars for both the model and the Auriga
halo.}
    \label{fig:stellar_halo_phase_space_contours}
\end{figure*}

The true underlying gravitational potential for each halo is calculated directly from the simulations snapshot. The total gravitational potential for each halo is the sum of the potentials generated by the stars, gas, and DM. The density distribution of the spherical components (i.e, DM and stellar haloes) have been modeled using a multipole expansion as implemented in \texttt{AGAMA}. In this scheme, we assume that the potential is a sum of contributions from various multipole moments by expressing it as the product of spherical harmonics and an arbitrary function of radius, $\Phi(r,\theta,\phi) = \sum_{l,m} \Phi_{l,m}(r)\mathrm{Y}_{l}^{m}(\theta,\phi)$. Each term's radial dependence is described by a quintic spline defined by a series of grid nodes spread in log $r$. The order of the angular expansion, $l_{\mathrm{max}}$, is determined by the shape of the density profile. The density distributions for the flattened/disky components (i.e., gas and in-situ stars) are calculated through an azimuthal harmonic expansion. The potential is assumed to be a sum of Fourier terms in the azimuthal angle, i.e., $\sin(m_{i}\phi)$, $\cos(m_{i}\phi)$. The coefficients of each term are interpolated on a 2d quintic spline in the ($R,z$) plane. As before, the order of the angular expansion, $m_{\mathrm{max}}$ is determined by the density profile shape. The full technical details of these implementations are given in the \texttt{AGAMA} documentation \citep{Vasiliev2018_Agamaext}. The density then follows from the Poisson equation.

\par Figure \ref{fig:totdens_Auriga} shows the comparison between the total spherically averaged density profiles of the best-fit model and the Auriga haloes. They follow an approximate single power-law profile and agree well. Figure \ref{fig:encmass_tot_Auriga} compares the total true enclosed mass profiles with the best-fit profiles. Similarly to the density profiles, due to the presence of substructure and other sources of anisotropy in the haloes, we spherically average within thin spherical shells when computing the profiles. Again the agreement is excellent.

\par The best-fit DM density profiles in Figure \ref{fig:dens_DM_Auriga} and enclosed DM mass profiles in Figure \ref{fig:encmass_DM_Auriga} show a good overall fit to the true DM density and mass profiles, particularly in the outer halo regions. Furthermore, the dynamical model recovers a range of axis ratios for the DM haloes: $q_{\mathrm{Au21}} = 0.93$, $q_{\mathrm{Au23}} = 0.60$, and $q_{\mathrm{Au24}} = 0.72$. The higher DM halo flattening of haloes 23 and 24 could be caused by what appears to be an ex-situ disk, which would lead to a more significant baryonic effect in the more central regions (see also \cite{Read+2009}).

\subsection{The stellar halo distribution function}
Here we present the results for the fitting to the stellar halo distribution function. 

Figure \ref{fig:stellar_halo_phase_space_contours} compares the joint distribution of pairs of ($\mathbf{x},\mathbf{v}$) phase-space coordinates. The colour-filled contours illustrate stellar halo phase-space distribution recovered by our best-fit model, while the black contours show the true phase-space distribution of the Auriga 24 halo. It can be seen that, overall, there is good agreement between the two distributions. The distribution of the density and velocity ellipsoids are discussed in more detail below.

\subsubsection{Number density}
\begin{figure*}
    \includegraphics[width=\textwidth]{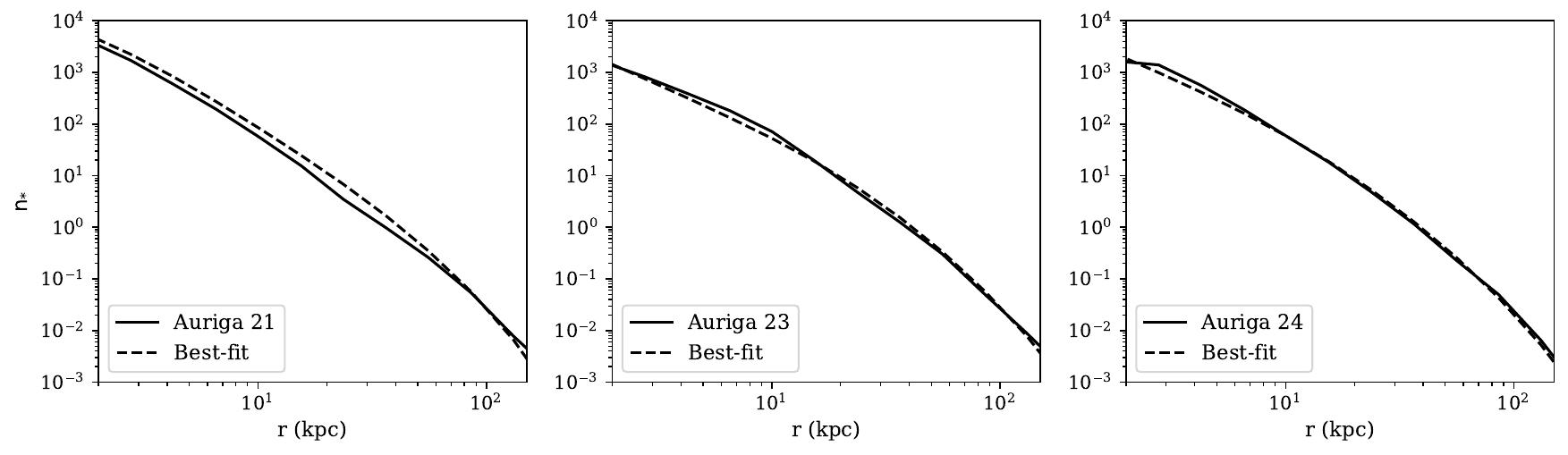}
    \caption{Number density profiles of the true Auriga stellar halos and the best-fit models.}
    \label{fig:ndens_shalo_Auriga}
\end{figure*}

Figure \ref{fig:ndens_shalo_Auriga} shows the number density profiles for the stars in the best-fit models and in the investigated Auriga haloes. The number density profiles of the Auriga stellar haloes have been calculated by binning the accreted stellar particles from the simulation snapshot into radial bins and computing the corresponding number density in each bin:
\begin{align} 
\begin{split}
   & n_{*}(\bar{r}_{i}) = \frac{N_{*,i}}{\Delta V_{i}},  \\ 
   & \text{where}\,\, \Delta V = \frac{4\pi}{3}(r_{i+1}^3-r_{i}^3), \\
\end{split}
\end{align}
with $N_{*,i}$ being the number of stars in the radial bin $i$ with edges $r_{i}$ and $r_{i+1}$.
To obtain the number density for the best-fit model, we sampled a number of halo stars equal to the number of accreted stellar particles in the corresponding Auriga halo snapshot and proceeded to calculate the number density profile as described above.

It can be seen that there is very good agreement between the Auriga and best-fit profiles, with both following a power-law profile with similar inner and outer slopes.

\subsubsection{Velocity distributions} \label{subsubsec:vrot_mean_Auriga}
\begin{figure*}
    \includegraphics[width=\textwidth]{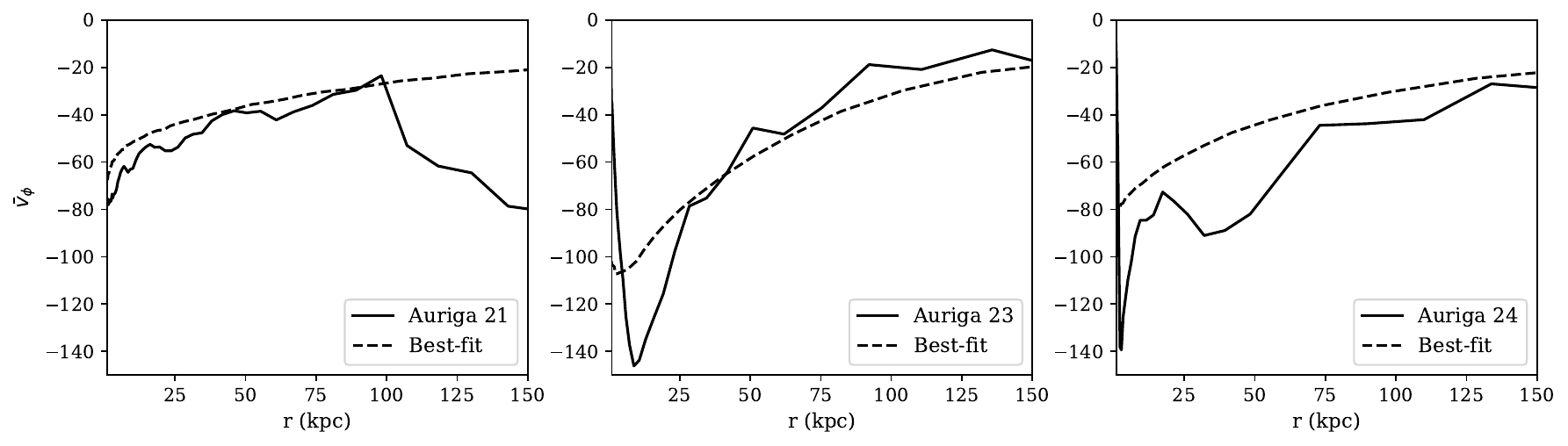}
    \caption{Mean $\mathrm{v}_{\phi}$ against radius for all Auriga stellar halos. Both the true stellar halos and best-fit models show an overall net, counter-clockwise rotation.}
    \label{fig:vphi_mean_shalo_Auriga}
\end{figure*}

Figure \ref{fig:vphi_mean_shalo_Auriga} compares the mean rotational velocity $\overline{\mathrm{v}}_{\phi}$ vs. radius between the true Auriga halos and best-fit models. The best-fit $\overline{\mathrm{v}}_{\phi}$ profile has been obtained by computing the first moment of the DF

\begin{equation}
\mathbf{\bar{v}} = \frac{1}{n_{*}}  \iiint\mathrm{d^{3}v}\mathbf{v}f(\mathbf{J}),
\end{equation}

where $n_{*}$ is the zeroth moment of the DF.
The true $\bar{\mathrm{v}}_{\phi}$ profile has been obtained by binning the accreted Auriga stars into radial bins and computing the average $\mathrm{v}_{\phi}$ in each bin. It can be seen that for both the true Auriga stellar halos, and recovered best-fit models, there is a net rotation of the component. While the equilibrium best-fit model predicts a smooth profile, the undulations in the true Auriga 24 profile are likely to be related to substructure and the number of stars per bin. Stellar halos 23 and 24 show a much larger net rotational velocity ($\bar{\mathrm{v}}_{\phi} \approx 140$ km/s) in the inner regions compared to halo 21 ($\bar{\mathrm{v}}_{\phi} \approx 80$ km/s). Furthermore, halo 21 shows a sharp increase in rotational velocity at about $r \approx 100$ kpc to values of $\bar{\mathrm{v}}_{\phi} \approx -80 $ km/s.

\subsubsection{Stellar halo velocity anisotropy}
\begin{figure*}
    \centering
    \includegraphics[width=0.95\textwidth]{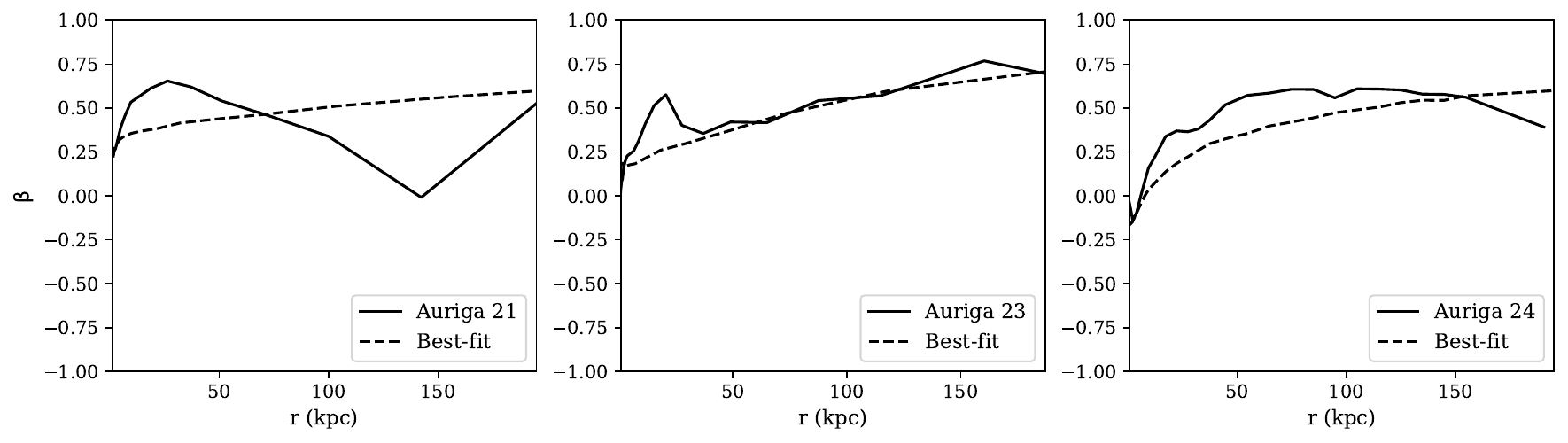}
    \caption{Anisotropy profiles of the stellar haloes in Auriga. They follow a radial bias which increases with radius. Halo 21 shows a sharp tangential dip between $r \approx 100-200$ kpc.}
    \label{fig:anisotropy_Auriga}
\end{figure*}

Figure \ref{fig:anisotropy_Auriga} shows the best-fit and true Auriga stellar haloes' anisotropy profiles for all of the three investigated haloes. The best-fit models predict anisotropy profiles rising sharply from tangential to radial (asymptoting at $\beta \approx 0.5$). The Auriga haloes show the same general trend, but also display a series of tangential dips and radial peaks in regions where the equilibrium best-fit model is smooth. These departures from the equilibrium anisotropy profiles are expected to arise as a result of more recent merger events that have not had time to phase mix. The data-derived anisotropy profiles should therefore hold clues about the more recent accretion history of the haloes compared to the profiles obtained through equilibrium modelling (i.e., the best-fit results). The latter should reflect merger events further back in time. This is further discussed in section \ref{sec:discussion}.
\par Halo 21 shows a very pronounced tangential dip between $r \approx 100 - 200$ kpc. This feature seems to overlap with the region of high $\bar{\mathrm{v}}_{\phi}$ velocity discussed in subsection \ref{subsubsec:vrot_mean_Auriga}.   However, this is probably a short-lived structure, that will phase mix quickly, why it isn't recovered by our equilibrium model.

\subsubsection{Stellar halo flattening} \label{subsubsec:flattening_stellarhalo_Auriga}
\begin{figure*}
    \includegraphics[width=\textwidth]{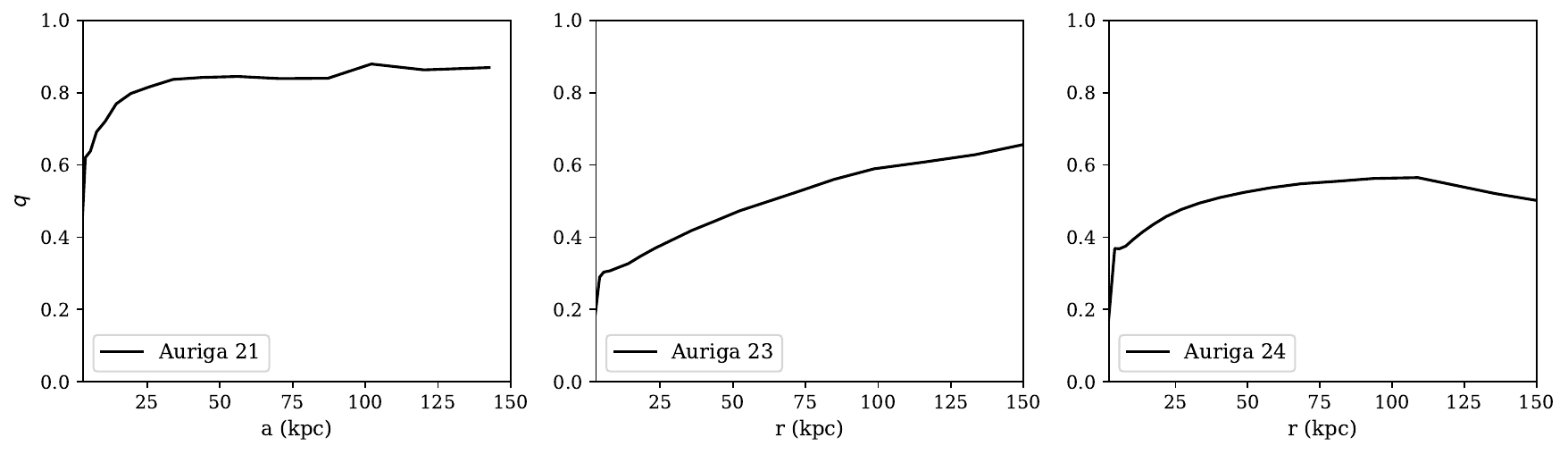}
    \caption{Stellar halo flattening profiles of the Auriga stellar halos as predicted by the best-fit model.}
    \label{fig:flattening_shalo_Auriga}
\end{figure*}

\begin{figure}
\centering
\begin{subfigure}{0.48\textwidth}
    \includegraphics[width=1\linewidth]{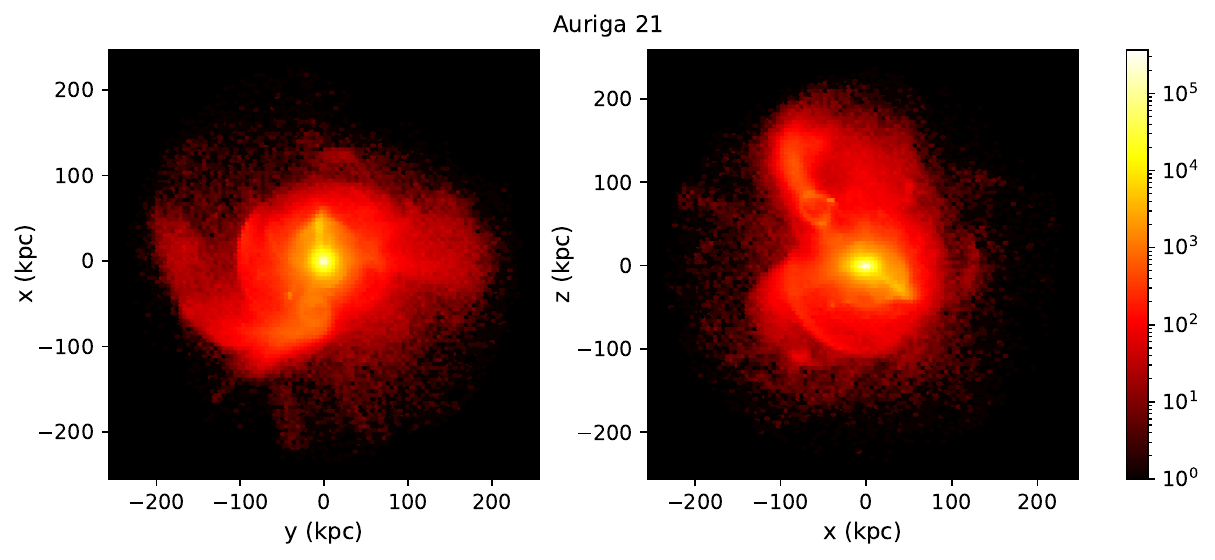}
    \caption{}
    \label{fig:exsitu_disk_Au21}
\end{subfigure}

\begin{subfigure}{0.48\textwidth}
    \includegraphics[width=1\linewidth]{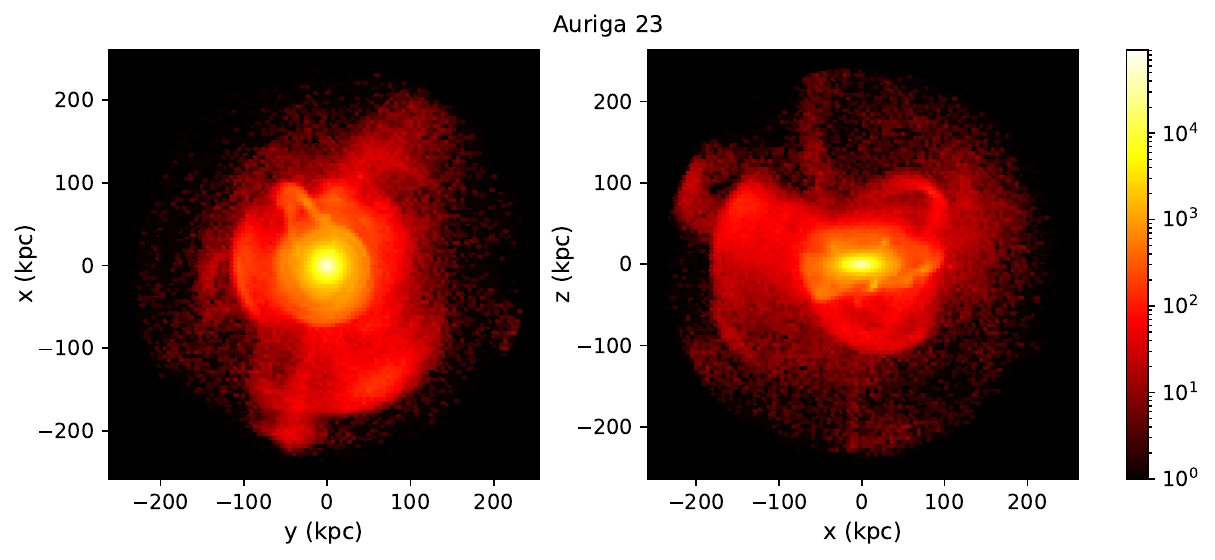}
    \caption{}
    \label{fig:exsitu_disk_Au23}
\end{subfigure}

\begin{subfigure}{0.48\textwidth}
    \includegraphics[width=1\linewidth]{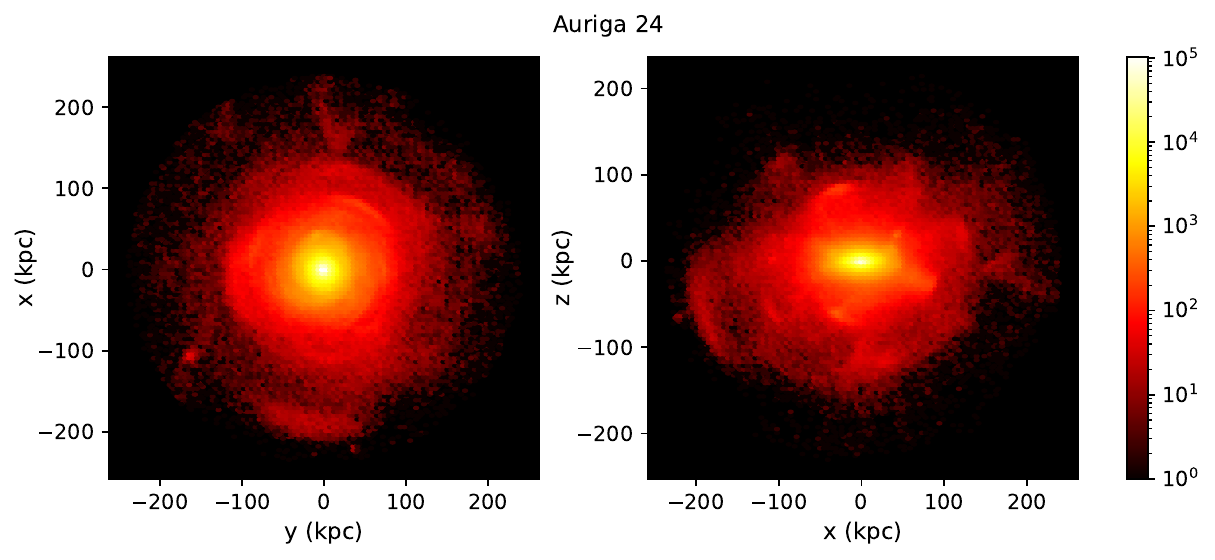}  
    \caption{}
    \label{fig:exsitu_disk_Au24}
\end{subfigure}

\caption{Hexbin projections of the stellar haloes (i.e., accreted stars only) of the three Auriga haloes investigated in the $x$-$y$ and $x$-$z$ planes. A disk-like structure is noticeable for haloes 23 and 24.}
\label{fig:exsitu_disk_Au23_Au24}

\end{figure}

Figure \ref{fig:flattening_shalo_Auriga} shows the axis ratio, $q$, profile against elliptical radius, $a$, as predicted by the best-fit model. The profiles have been computed in the same way as described in section \ref{subsec:mock_shalo_flattening}. Halo 21 has a relatively uniform axis ratio of $q \approx 0.8 $ throughout, which indicates a more spherical halo. Halo 23 appears to be more flattened throughout, with $q \approx 0.2$ in the inner region and uniformly increasing to $q \approx 0.6$ in the outer halo. Halo 24's profile indicates a stellar halo that is also highly flattened, with the inner halo more flattened ($q \approx 0.3$) than the outer halo ($q \approx 0.5$). 

The flattening of the stellar halos could have been influenced by different factors such as the galaxy formation history and dark matter distribution. The high flattening of stellar haloes 23 and 24 at low radii ($q \approx 0.2$) can also be expected to arise due to the much larger rotational velocity in these regions $\bar{\mathrm{v}}_{\phi} \approx 140$ km/s (see Figure \ref{fig:vphi_mean_shalo_Auriga}). On the other hand, halo 21 shows a much more spherical inner stellar halo, while also displaying a lower rotational velocity in this region $\bar{\mathrm{v}}_{\phi} \approx 80 $ km/s.

\par We believe the high central rotational velocity and high flattening of the haloes 23 and 24 could be due to a possible ex-situ disk. Furthermore, Figure \ref{fig:exsitu_disk_Au23_Au24} shows a disk-like structure in the haloes' hexbin plots. \cite{Gomez+2017} discuss such an ex-situ disk for halo 24, but not for halo 23. It is also interesting to note that haloes 23 and 24 also display a more flattened DM halo (see subsection \ref{subsec:mass_distribution_Auriga}).

\section{Discussion} \label{sec:discussion}
In this section, we discuss the suitability of action-based double-power law DF to describe M31-like stellar halos, the ability of our model to recover the mass profiles of the galaxy (including the DM), as well as discuss the stellar halo anisotropy profiles from a merger history perspective.

\begin{figure*}
\begin{subfigure}{0.95\textwidth}
    \includegraphics[width=.95\linewidth]{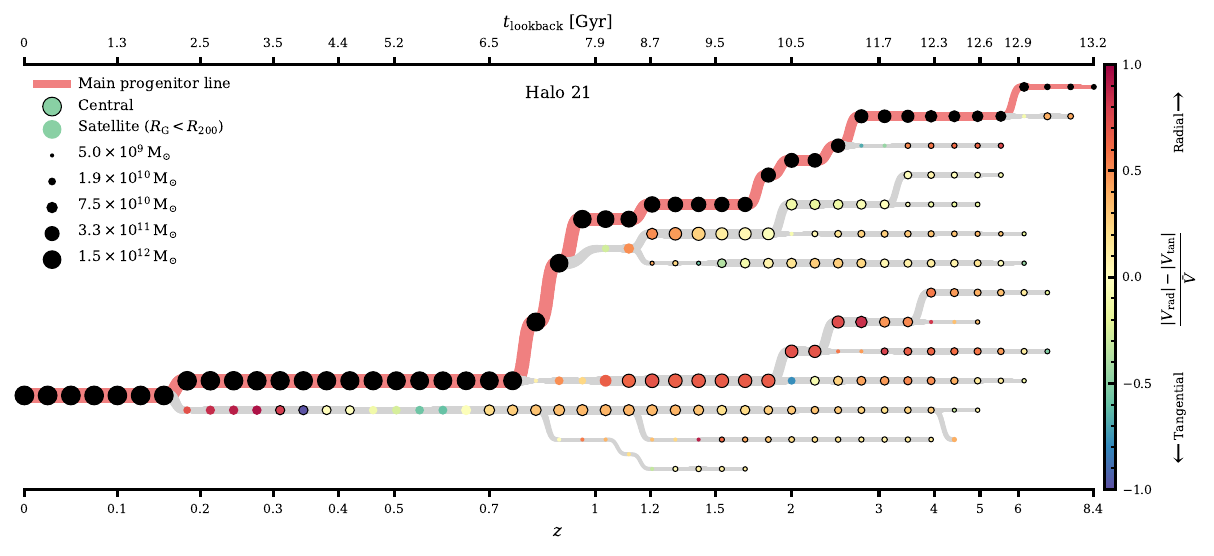}
    \caption{Merger tree of Auriga halo 21.}
    \label{fig:merger_tree_Au21}
\end{subfigure}
\begin{subfigure}{0.95\textwidth}
    \includegraphics[width=.95\linewidth]{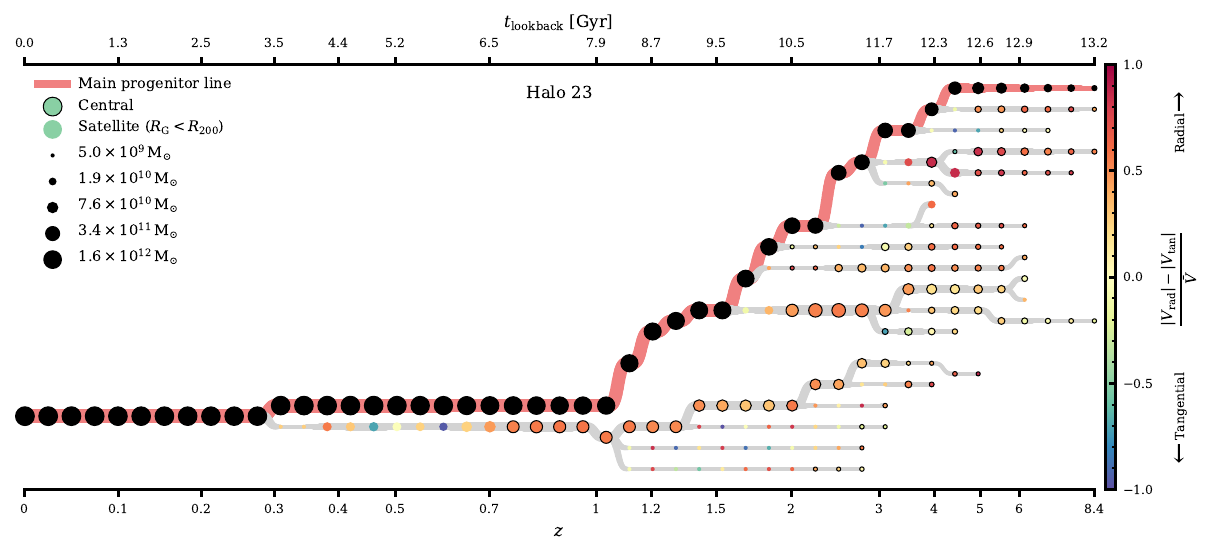}  
    \caption{Merger tree of Auriga halo 23.}
    \label{fig:merger_tree_Au23}
\end{subfigure}
\end{figure*}

\begin{figure*} \ContinuedFloat
\begin{subfigure}{0.95\textwidth}
    \includegraphics[width=.95\linewidth]{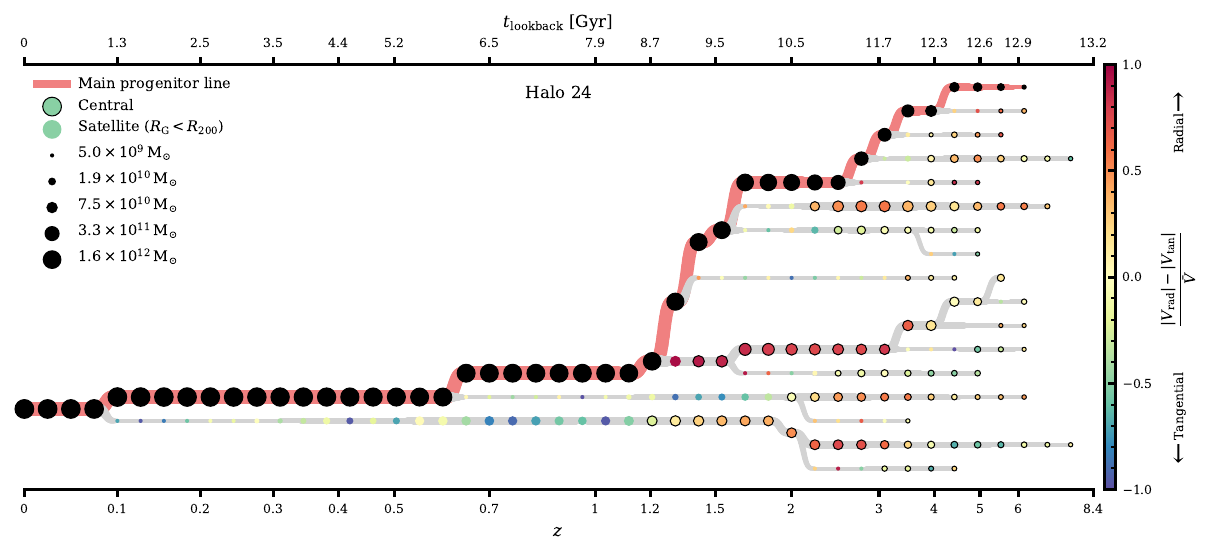}  
    \caption{Merger tree of Auriga halo 24.}
    \label{fig:merger_tree_Au24}
\end{subfigure}

\caption{Merger trees of the 3 investigated halos (a) Auriga 21, (b) Auriga 23, and (c) Auriga 24. The plot should be read from right to left. Each node corresponds to a halo with the connecting lines indicating descendent to the left and progenitors to the right. The main progenitor is marked in pink. The color of the nodes indicate how radial or tangential an orbit is, while the size correlates with the halo mass. The velocities are relative to the host halo.}
\label{fig:merger_trees_Auriga}
\end{figure*}

\subsection{Double-power law DFs for M31-like stellar haloes}
In this work, we have assumed a stellar halo described by a double-power law DF as introduced by \citet{Posti+2015_sphDF}. The DF predicts a double power law in the stellar density profile, which provides an excellent fit to the three Auriga stellar haloes modelled here. The profiles of rotation velocity and velocity anisotropy for the simulated stellar haloes have a number of peaks and troughs as a result of non-phase-mixed material that has not been removed. The DF however predicts smooth profiles that recover the global structure of these profiles.

\subsection{Recovery of the total and dark matter mass and density profiles, and the assumption of dynamical equilibrium}
 
Figure \ref{fig:fracdiff_totencmass_Auriga} shows the absolute fractional difference between the total enclosed mass predicted by the best-fit model and the true total enclosed mass of the haloes. Even though the degree of bias between the best fit and the true values varies from halo to halo, the systematics associated with the total mass of the haloes are below $\approx 20\%$.

\par There is a greater degree of discrepancy between the Auriga dark matter density and enclosed dark matter mass profiles and the models, particularly in the inner regions. This may be a consequence of the reduced freedom in the inner region of the mass model, as we only allow the total masses of the bulge and disk components to vary during the fitting rather than their shapes. From a computational expense point of view and, since our goal is to focus to constrain the outer dark matter content and the distribution function of the stellar halo, this is a reasonable assumption to make.

The overall agreement in the total mass profiles and the DF moment profiles suggests that dynamical equilibrium is a reasonable assumption to make, even in these relatively disturbed haloes. There has been previous work investigating the effects of the equilibrium assumption in mass modelling in stellar haloes of simulations \citep[e.g][]{Sanderson+2017, Eadie+2018, Wang+2018}, which have overall found a possible lower boundary of $20\%$ in the accuracy of the total mass determination using stellar halo stars, with some haloes providing very inaccurate fits (for certain haloes with exceptional evolutions in \citealp{Eadie+2018} ). Our method may have performed better than past studies as the model is specified in action space. Non-phase-mixed substructures from low mass accretion events should conserve actions and therefore still look smooth in action space, even if they are structured in phase space. This should result in narrower confidence intervals as there is less noise in action space. The substructure may also not be symmetric about the smooth model in phase space, leading to biased mass estimates for models in phase space.

\subsection{NFW profiles for dark matter haloes of M31-like systems}

Overall, the NFW profiles provides an accurate description of the dark matter haloes of the Auriga haloes, with the exception of halo 21, which has a steeper inner density slope. In reality, there are different factors that could affect the DM density profile shape, making it cuspier or more cored through various feedback mechanisms.

In regions with high concentrations of baryonic matter, such as central galactic regions, the pull exerted by the baryons can cause the DM to be more concentrated through adiabatic contraction \citep[e.g][]{Cole_Binney2017}, which could explain the higher inner DM density in halo 21. The specific merger history of galaxies also has an effect on the shape and density of the DM halo, as the DM will respond to the baryonic effects introduced by the newly acquired mass during a merger. At the same time, inaccuracies at the center of the model can be due to the fact that we fixed the scale radii of both the bulge and disk when fitting the halos. Furthermore, the resolution limit of the simulations could also play a role.

\begin{figure*}
    \centering
    \includegraphics[width=0.95\textwidth]{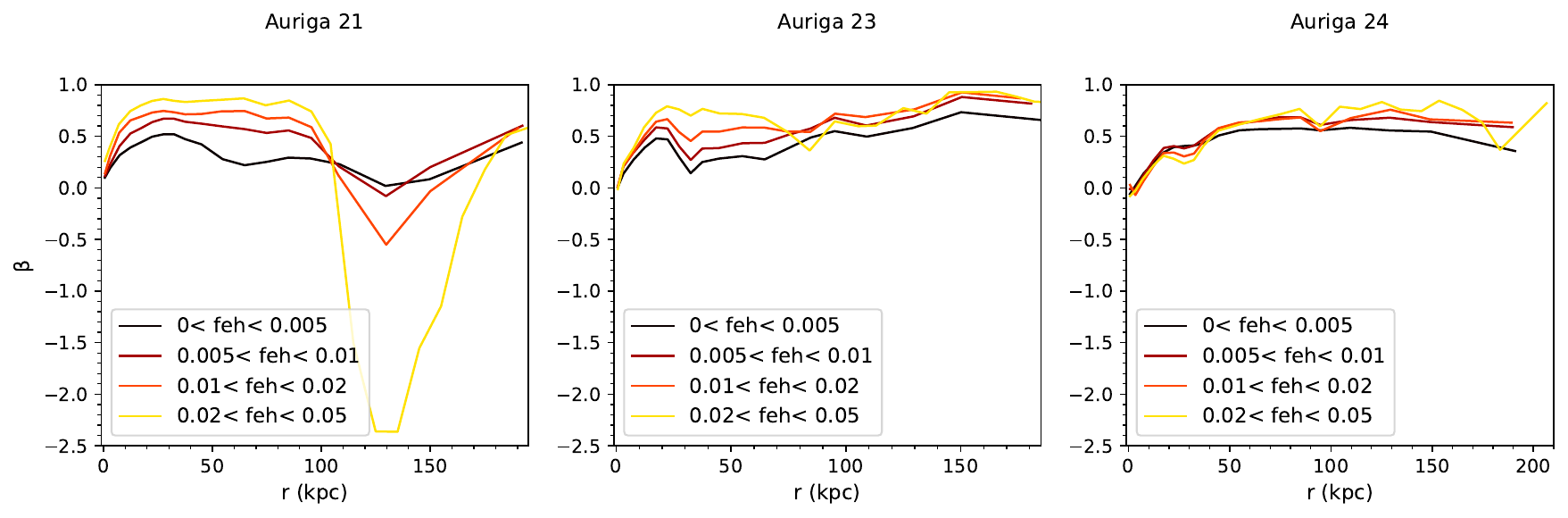}
    \caption{Variation of anisotropy profiles in Auriga with metallicity.}
    \label{fig:anisotropy_Auriga_metallicity}
\end{figure*}

\subsection{Anisotropy profiles of stellar haloes and their merger histories} {\label{subsec:discussion_anisotropy}}

Galaxy formation models predict radial anisotropy that increases with radius: near isotropy near the center vs radial bias in the outskirts \citep{Amorisco2017}. Our equilibrium model does reproduce these results. However, while the Auriga haloes also follow this overall trend, their anisotropy profiles show a series of undulations. These are expected to be correlated with the merger history of the galaxy, specifically with merger relics that have not fully phase mixed yet.

\par Figure \ref{fig:merger_trees_Auriga} shows the merger trees for the investigated Auriga halos. Reading the plot right to left, we go from larger lookback times (i.e. closer to the Big Bang) to the present day (at $t_{\mathrm{lookback}}=0, z = 0$). Each node corresponds to a halo, while the connecting lines indicate descendants (to the left) and progenitors (to the right). The main progenitor is marked in pink. The color of the nodes indicates how radial or tangential an orbit is, while the size correlates with the mass of the halo.

\par Halo 21 shows a big tangential dip in $\beta$ ranging between $r\approx100-190$ kpc. Haloes 23 and 24 follow the general trend of the equilibrium profile predicted by the best-fit model, with less dramatic dips.  Figure \ref{fig:merger_tree_Au21} shows that halo 21 has a merger that skims through the $\mathrm{R}_{200}$ radius just after $z=0.7$, like a skipping stone. This could be responsible for generating the large anisotropy dip in the outer regions of halo 21, as seen in Figure \ref{fig:anisotropy_Auriga}. The dip in the halo 21 anisotropy profile is correlated with a relatively metal-rich component (see Figure \ref{fig:anisotropy_Auriga_metallicity}), which supports the hypothesis of it coming from a recently accreted component. The anisotropic structure this merger introduces is probably short-lived and should phase-mix after a few dynamical timescales. 

\par \cite{Gomez+2017} claim that the most significant progenitor which contributed to the ex-situ disk investigated for halo 24 first crosses the virial radius at $t \approx 8.6 $ Gyr. This can be seen in Figure \ref{fig:merger_tree_Au24} as a merger that spends several Gyr spiraling around and around on a very tangential trajectory. The stellar debris from this particular merger is probably what created the ex-situ disk examined in \cite{Gomez+2017}.

\par The model is not constructed to take into account the cosmological context (i.e mergers, interactions with other galaxies, etc.) and assumes dynamical equilibrium. Therefore our equilibrium model will not be able to reproduce these dips and structures seen in the Auriga anisotropy profiles. However, our dynamical model can provide information about the older accretion and merger events, which have already phase-mixed. Furthermore, our model can provide hints of a possible ex-situ disk by examining the rotational velocity profiles and flattening profiles as discussed in Subsections \ref{subsubsec:vrot_mean_Auriga} and \ref{subsubsec:flattening_stellarhalo_Auriga}.

The models fit to haloes 21 and 23 show a greater degree of radial anisotropy than halo 24. From the merger trees information (see Figure \ref{fig:merger_trees_Auriga}), it can be seen that haloes 21 and 23 have experienced a large number of (major) mergers in a short span of time: $\approx$ 6.5 - 7.9 Gyr ago for halo 21, and $\approx$ 9.5 - 11.7 Gyr ago for halo 23. Halo 24, on the other hand, shows a more spread-out distribution of major merger events through time. A possible explanation for this difference is that, in the case of haloes 21 and 23, the progenitors of the early massive mergers have quickly sunk into the potential of the host at low radii through the more significant effects of dynamical friction. This could have led to a redistribution of orbital energies and angular momenta. In the process, stars that were initially on more tangential or random orbits can be scattered onto more radial orbits. 


\section{Conclusions} 
\label{sec:conclusion}
In this paper, we presented an action-based dynamical model designed to recover the mass distribution, DM content, and the stellar halo distribution function of M31-like halos in the Auriga simulation suite. Our model assumes dynamical equilibrium and comprises a galactic potential for the bulge, disk, and dark matter halo and an action-based DF for spherical components (based on \cite{Posti+2015_sphDF}) to represent the stellar halo. We fit for the potential and DF using i.) a mock data set generated from the model itself; ii) accreted stars only from three haloes in the Auriga simulations.

Our best-fit models provide a very good fit for the total mass and DM distribution of the Auriga haloes, while the recovered DFs prove to be an excellent description for the investigated stellar halos. The total mass is recovered with a fractional difference with a maximum of $20 \%$ (see Figure \ref{fig:fracdiff_totencmass_Auriga}), while the DM halo mass profile shows a slightly higher fractional difference in the inner regions, but still provides a good fit (see Figure \ref{fig:fracdiff_DMencmass_Auriga}). This is likely a result of the rigidity of the disk and bulge contributions to the total mass profiles. The overall agreement suggests that dynamical equilibrium is a reasonable assumption to make, even in the phase of relatively disturbed stellar haloes.

\par The anisotropy profile of the equilibrium dynamical model can shed light on the past merger history of the galaxy. The degree of radial anisotropy may reflect the mass ratio of early accretion events. However, to learn about the more recent merger history, the data-derived anisotropy profiles, in particular their variation with metallicity may be more informative.

In future work, we will apply the dynamical equilibrium model to the real M31 stellar halo in order to constrain its properties and get further insights into its unique formation pathway. Therefore, understanding the systematics and shortcomings of our model will enable us to make a more informed judgment on this future work.

\section*{Acknowledgements}
PG would like to thank the GLEAM team at the University of Surrey for the constructive conversations which took place while writing the paper. PG would also like to thank Dr. Eugene Vasiliev for the AGAMA software which was used extensively in this work. PD is supported by a UKRI Future Leaders Fellowship (grant reference MR/S032223/1). RG acknowledges support from an STFC Ernest Rutherford Fellowship (ST/W003643/1). MO acknowledges funding from the European Research Council (ERC) under the European Union’s Horizon 2020 research and innovation programme (grant agreement No. 852839). 

\section*{Data availability}
The Auriga simulations and Agama software are available publicly. The mock galaxy tests and code are available on request.

\bibliographystyle{mnras}
\bibliography{main} 






\end{document}